\documentclass[aps,floatfix,superscriptaddress,longbibliography,prl,reprint]{revtex4-2}
\usepackage{graphicx} % Required for inserting images
\usepackage[utf8]{inputenc}
\usepackage{natbib}
\usepackage[normalem]{ulem}
\usepackage{xcolor}
\UseRawInputEncoding

\usepackage[tbtags]{amsmath}
\usepackage{hyperref}
\hypersetup{colorlinks,allcolors=blue}
\usepackage{amssymb}
\usepackage{gensymb}
\usepackage{float}
\usepackage{amsmath}
\usepackage{tabularx,graphicx}
\usepackage{epstopdf}
\usepackage{latexsym}
\usepackage{color, colortbl}
\usepackage{psfrag}
\usepackage{bbm}
\usepackage{bm}
\usepackage{titlesec}
\usepackage{dsfont}
\usepackage{feynmp}
\usepackage{slashed}
\usepackage{multirow}
\usepackage{soul} % strikethrough via \sout{...} 
\usepackage{braket}
\usepackage{pgfplots}
\usepackage{siunitx}

 \usepackage{commath}
 \usepackage{mathtools}
\usepackage{listings}
 \usepackage{comment}
 \usepackage{wrapfig}
 \usepackage{subfiles}
 \usepackage{multirow}
\usetikzlibrary{plotmarks}

\def\dens{{n}}

\begin{document}
\title{Electrostatically stabilized surface flat bands in rhombohedral graphite at zero displacement field}

\author{Kry\v{s}tof Kol\'a\v{r}}
\affiliation{Department of Applied Physics, Aalto University School of Science, FI-00076 Aalto, Finland}
\author{Andrea F. Young}
\email{andrea@physics.ucsb.edu}
\affiliation{Department of Physics, University of California at Santa Barbara, Santa Barbara CA 93106, USA}
\author{Cyprian Lewandowski}
\email{clewandowski@fsu.edu}
\affiliation{National High Magnetic Field Laboratory, Tallahassee, FL 32310, USA}
\affiliation{Department of Physics, Florida State University, Tallahassee, FL 32306, USA}
\date{\today}

\begin{abstract}
Rhombohedral (ABC-stacked) multilayer graphene hosts interaction-driven phases enabled by surface flat bands at large displacement fields. In thick flakes, however, strong screening suppresses internal electric fields, raising the question of whether a flat-band regime is accessible within the same experimental paradigm. Here, we show that self-consistent, nonlinear electrostatics provides a robust alternative mechanism: even in the absence of a displacement field, a nonuniform near-surface potential flattens the surface-band dispersion and enhances the density of states. In the strong-coupling limit, electrostatics drives the system toward uniform half-filling at each momentum, yielding an asymptotically flat surface band without any gating. At realistic interaction strengths, surface-band flatness is tuned by the proximal gate, with maximal flatness achieved at hole doping when the band is empty. Combining analytic arguments with fully self-consistent calculations in a realistic model, we map the resulting low-field regime and connect to finite $N\!\sim\! 6-15$ layered samples, providing a framework for analyzing the symmetry-broken phases observed in these systems. Our results motivate future experiments in large-$N$ devices and establish a low-field regime for exploring electrostatically induced flat-band physics.
\end{abstract}

\maketitle
\begin{figure}[!t]
    \centering
    \includegraphics[width=0.85\columnwidth]{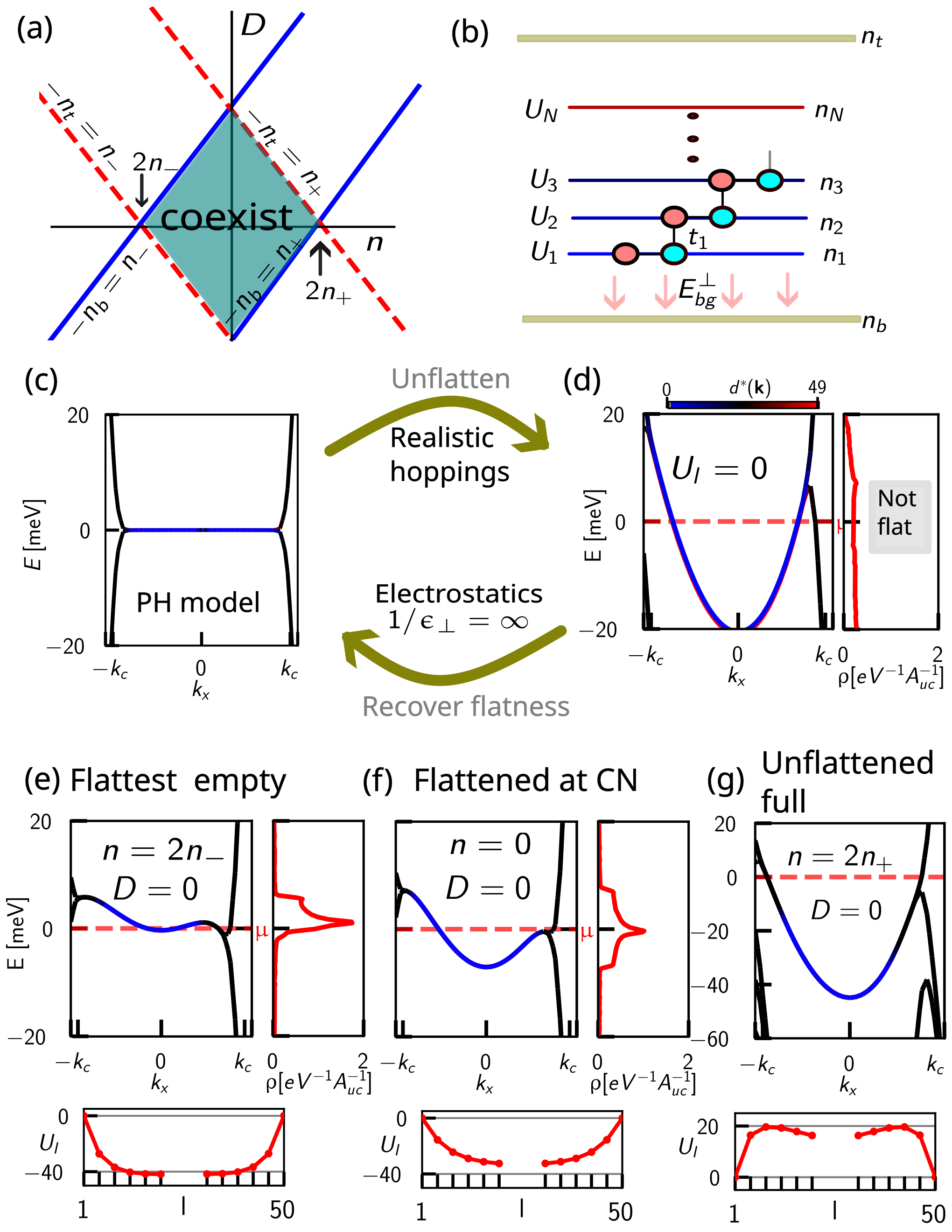} %Non-wild version figone.png
    \caption{ 
(a) Schematic phase diagram of large $N$ RNG in the $n$-$D$ plane. (b) Device schematics, showing the layer and gate electron densities. The gate electric field is denoted as $E^\text{bg}_\perp$.
(c) Band structure for an idealized $N=50$ RNG model at charge neutrality.
    (d) Band structure and density of states (DOS) for a realistic model at $N=50$ for zero layer potential, $U_l=0$.
    (e-g) Upper panel shows the self-consistent band structure and DOS at $D=0$ and: (e) negative $n \approx 2n_-$, (f) $n=0$, and (g) close to full filling, $n \approx 2n_+$. 
    Lower panel shows the corresponding layer potentials $U_l$.}
    \label{fig:figone}
\end{figure}

Rhombohedral (ABC-stacked) $N$-layer graphene (RNG) hosts surface-polarized bands with an exceptionally large density of states and has become a prime venue for exploring interaction-driven phases \cite{mishchenkoShiElectronicPhaseSeparation2020,falkoSlizovskiyFilmsRhombohedralGraphite2019,zhouZhangCorrelatedTopologicalFlat2024,yinZhangLayerdependentEvolutionElectronic2025,PhysRevB.81.161403,PhysRevB.97.245421,Zhou2021,Zhou2021-b,Liu2023,Auerbach2025,Han2023,Guo2025arXiv_ThickRhombohedral_SC,Kumar2025arXiv_DualSurface_SC,Deng2026arXiv_Hexalayer_Semimetal,2025arXiv250405129Q,Zhou2024,juLuExtendedQuantumAnomalous2025,lauMyhroLargeTunableIntrinsic2018,nemes-inczeHagymasiObservationCompetingCorrelated2022,juHanCorrelatedInsulatorChern2024,PhysRevX.15.011045,youngChoiSuperconductivityQuantizedAnomalous2025,Aronson2025PRX_DisplacementControlled_FCI_CDW,Han2025,juSeoFamilyUnconventionalSuperconductivities2025a,liXuSignaturesUnconventionalSuperconductivity2026}. Most works have focused on few-layer devices at large displacement fields, where a \emph{single} surface band is both isolated and strongly flattened, enabling, amongst others, the observation of fractional Chern physics and chiral superconductivity \cite{Lu2024,PhysRevX.15.011045,youngChoiSuperconductivityQuantizedAnomalous2025,Aronson2025PRX_DisplacementControlled_FCI_CDW,Han2025,juSeoFamilyUnconventionalSuperconductivities2025a,liXuSignaturesUnconventionalSuperconductivity2026}. More recently, experiments have also begun mapping a lower-displacement field in thicker samples and have found a semimetallic coexistence regime in the $(n,D)$ plane \cite{Guo2025arXiv_ThickRhombohedral_SC,Kumar2025arXiv_DualSurface_SC,Deng2026arXiv_Hexalayer_Semimetal,Zhou2024NatCommun_LayerPolarizedFM,huntSeifertIncreasingFlatnessSurface2024}. This regime shows as a characteristic ``diamond'' (Fig.~\ref{fig:figone}a) in a double-gated geometry (Fig.~\ref{fig:figone}b) where key experimental features follow lines of constant \emph{single-gate} voltage  - indicating the simultaneous involvement of carriers on both surfaces \cite{Kolar2026PRB_SingleGateTracking,Guo2025arXiv_ThickRhombohedral_SC,Kumar2025arXiv_DualSurface_SC,Deng2026arXiv_Hexalayer_Semimetal}. Motivated by this phenomenology and emerging experimental interest in thicker RNG samples, we study the complementary limit of large $N$ and small (or zero) $D$, where both surface bands can remain relevant but become increasingly \emph{electrostatically decoupled}.

A central open question we would like to address is what becomes of the ``flat-band'' physics in the $N\!\to\!\infty$
limit. Idealized chiral models predict a surface dispersion proportional to $|\mathbf{k}|^N$, leading to increasing
``flatness'' with $N$
\cite{Guinea2006PRB_GrapheneStacks,Koshino2009PRB_TrigonalWarping,Zhang2010PRB_ABCTrilayerBandStructure,volovikHeikkilaDimensionalCrossoverTopological2011,heikkilaKopninHightemperatureSurfaceSuperconductivity2013,PhysRevB.77.155416}. In realistic
band structures, remote hoppings qualitatively change the long-wavelength behavior
\cite{Koshino2009PRB_TrigonalWarping,Zhang2010PRB_ABCTrilayerBandStructure,McCann2013RoPP_BilayerReview,jungParkTopologicalFlatBands2023}.
Such terms generate a conventional quadratic surface-band dispersion with positive effective mass
even as $N\!\to\!\infty$. This outcome challenges the naive expectation of an asymptotically flat band (see
Fig.~\ref{fig:figone}c,d for the bands for $N=50$)
\cite{Koshino2009PRB_TrigonalWarping,Zhang2010PRB_ABCTrilayerBandStructure}. For small $N{\sim} 4\!-\!6$, a high
displacement field $D$, typically modeled by a linear potential drop between layers, restores an effective flat band
\cite{Zhang2010PRB_ABCTrilayerBandStructure,serbynGhazaryanMultilayerGraphenesPlatform2023,McCann2013RoPP_BilayerReview}.
However, for large $N$, a linear potential is not an adequate model for the internal potential profile~\cite{koshinoKoshinoInterlayerScreeningEffect2010,PhysRevB.80.195401,PhysRevB.81.115432,Kolar2026PRB_SingleGateTracking,Slizovskiy2021,PhysRevB.81.161403}
and generally does not produce an isolated flat band. In this work, we show that, in the
large-$N$ limit ($N\gtrsim6$), accurately accounting for nonlinear electrostatics can instead yield a flat surface band
at $D=0$ (Fig.~\ref{fig:figone}e,f). 

Our main result is a simple electrostatic mechanism that flattens the quadratically-dispersing (Fig.~\ref{fig:figone}d) surface band in rhombohedral graphite. At charge neutrality and when the out-of-plane interaction is strong ($1/\epsilon_\perp\!\to\!\infty$), the electrostatic energy is lowest if every layer stays neutral. This happens when each surface-band momentum state $\mathbf{k}$ is half-filled, which can only occur if the mean-field particle dispersion is flat. Although this argument assumes in-plane isotropy, a similar principle works for realistic, trigonally warped dispersions in RNG. 

While the proposed flattening mechanism requires no gating and holds at the charge neutrality point, electrostatic gating still provides a useful tuning knob in large-$N$ devices. The key distinction from thin stacks is, however, the screening: in a double-gated geometry (Fig.~\ref{fig:figone}b), the distant gate is almost fully screened from a given surface band \cite{Kolar2026PRB_SingleGateTracking}. Accordingly, instead of the usual experimental variables $n=-n_t-n_b$ and $D=-(n_t-n_b)/\epsilon_0$, the bottom surface, for example, is controlled primarily by $n_b$ with $n_t$ being largely irrelevant \cite{Kolar2026PRB_SingleGateTracking}. This observation lets us follow the bottom-surface (or top-surface) evolution along essentially any trajectory  in the $(n,D)$ plane as the corresponding $n_t$ (or $n_b$) coordinate is a spurious parameter; to specifically highlight the contrast with few-layer devices, we thus focus on the $D=0$ line.

From this viewpoint, the bottom gate sets both the surface filling and the inner electric field. As a result, the self-consistent potential is highly non-linear: the potential drops decay from the surface to bulk instead of being constant \cite{koshinoKoshinoInterlayerScreeningEffect2010,PhysRevB.80.195401,PhysRevB.81.115432,Kolar2026PRB_SingleGateTracking}. A potential-induced dispersion arises as
$\sum_l U_l W_l(\mathbf{k})$, where $W_l(\mathbf{k})$ is the normalized layer weight, 
given in the minimal chiral model as
\begin{equation}
\label{eq:wlpuremodel}
W_l(\mathbf{k})=(1-x)x^{l-1},\qquad
x=\left(\frac{v_F|\mathbf{k}|}{t_1}\right)^2,
\end{equation}
with $v_F$ the Dirac velocity and $t_1$ the dominant $B_l\!\to\!A_{l+1}$ hopping (Fig.~\ref{fig:figone}b) \cite{Zhang2010PRB_ABCTrilayerBandStructure}.
The $x=0$ ($\mathbf k=0$) state is completely polarized to the bottom layer, and states with higher values of $x$ becomes progressively 
more layer-delocalized. To quantify the polarization, we colorcode with the average layer index
$d^*(\mathbf{k})=\sum_l l\,W_l(\mathbf{k})$ in band structure plots.
A potential that decreases into the sample (a $U$-shape at $D=0$) counteracts the quadratic dispersion in Fig.~\ref{fig:figone}d by lowering the energy of the finite-$\mathbf{k}$ states, thereby flattening the band.
This type of potential forms at charge neutrality (Fig.~\ref{fig:figone}f, bottom) and 
naturally emerges in the infinite interaction limit ($1/\epsilon_\perp\!\to\!\infty$) discussed above.
Hole doping increases this flattening by steepening the $U$-shaped profile, due to the external bottom-gate electric field $E^\perp_{bg}=e n_b/(\epsilon_\perp\epsilon_0)$ (Fig.~\ref{fig:figone}b). In contrast, electron doping produces the opposite (inverted-$U$) profile, making the band more dispersive (Fig.~\ref{fig:figone}g). The rest of the paper develops this robust flattening mechanism in detail and connects it to finite $N$, focusing on the experimentally relevant case $N=13$.

{\bf Single-particle theory of RNG} To analyze the large-$N$ limit, we first consider an idealized model of rhombohedral graphite that is infinite along $z$, with a bulk Bloch Hamiltonian (analogous to an SSH chain~\cite{PhysRevLett.42.1698}) 
\begin{equation}
\label{eq:graphiteham}
H(\mathbf k ,k_z)=
    \begin{pmatrix}
0 &  v_F \overline{k}+ t_1 e^{-ik_z} \\
         v_F k + t_1 e^{ik_z}&0 \end{pmatrix},
\end{equation}
where $\mathbf{k}$ is the in-plane momentum, $k_z$ is the out-of-plane momentum, and $k=k_x+ik_y$.
This model hosts a zero-energy topological nodal spiral at $|\mathbf{k}|=t_1/v_F\equiv k_c$
\cite{volovikHeikkilaFlatBandsTopological2011,volovikHeikkilaDimensionalCrossoverTopological2011,balentsBurkovTopologicalNodalSemimetals2011}.
By bulk-boundary correspondence, terminating the system yields surface-localized zero-energy states for $|\mathbf{k}|\le k_c$, forming a surface flat band with
\begin{equation}
\label{eq:nc}
n_c=N_f\frac{\pi k_c^2}{4\pi^2}
= N_f\frac{1}{4\pi}\left(\frac{t_1}{v_F}\right)^2
=\SI{13.5d12}{cm^{-2}},
\end{equation}
states per unit area (with $N_f=4$ for spin-valley degeneracy). 

To understand the real-space vertical structure of the surface states, we explicitly solve for the surface modes by allowing complex $k_z$ and imposing $v_F k+t_1 e^{ik_z}=0$, i.e., $e^{ik_z}=-v_F k/t_1$. An unnormalized surface wavefunction for a semi-infinite stack then takes the form $(1, e^{ik_z}, e^{i2k_z},...)^T\otimes (1,0)^T$ (reflecting the layer and sub-lattice structure, where each component corresponds to a different layer). Physically, $e^{ik_z}$ is the layer-to-layer propagation amplitude, so the probability density decays as $|e^{ik_z}|^{2(l-1)}$.
Defining $x\equiv |e^{ik_z}|^2=\left(\frac{v_F|\mathbf{k}|}{t_1}\right)^2$, we obtain the normalized layer-resolved probability of Eq.~\eqref{eq:wlpuremodel}. 

A more realistic finite-$N$ description, which includes remote tunneling (electron hopping between distant layers) and layer potentials, uses the standard RNG multilayer Hamiltonian
\cite{falkoMcCannLandauLevelDegeneracyQuantum2006,peresNilssonElectronicPropertiesBilayer2008,McCann2013RoPP_BilayerReview}:
\begin{equation}
\label{eq:hamrhombo}
\hat H_{SP}(\mathbf k)=
    \begin{pmatrix}
        U_1 &  v_F \overline{k}&-v_4 \overline{k}&-v_3 k &0&-t_2/2& \\
         v_F k& U_1&t_1  &-v_4 \overline{k} &\\ 
         -v_4 k& t_1&U_2 &  v_F \overline{k}& \\
        -v_3  \overline{k}&-v_4 k& v_F k &U_2&  \\
        0& &&& \ddots\\
        -t_2/2& &&& &\ddots\\
    \end{pmatrix}.
\end{equation}
Here, $U_l$ are layer electrostatic potentials that will arise due to both external gates and self-consistent charge screening physics. The parameter $t_1$ is the dominant $B_l\to A_{l+1}$ hopping, parameters $v_3,v_4\ll v_F$ quantify nonlocal interlayer tunneling, and $t_2$ is the $A_l \to B_{l+2}$ hopping (parameter values are given in the Supplement~\cite{supplement}). The particle-hole (PH) symmetric (chiral) limit corresponds to $v_3=v_4=t_2=0$. In this limit, the low-energy surface states are located on the two undimerized outer sites and disperse as $E^{PH}_{\mathbf{k}}\propto |\mathbf{k}|^{N}$ (for $U_l=0$), highlighting the nominal ``band flattening'' for multilayer rhombohedral graphene, as shown in Fig.\ref{fig:figone}c. Crucially, the chiral model is an idealization, and the introduction of $v_4>0$ generates a quadratic surface-band dispersion (independent of layer number) even at $U_l=0$,
\begin{equation}
\label{eq:zeropotdispersion}
E_{\mathbf{k}}\big|_{U_l=0}\approx \frac{2v_4 v_F}{t_1}|\mathbf{k}|^2,
\end{equation}
as seen in Fig.~\ref{fig:figone}d for $N=50$.

{\bf Role of electrostatics in a gated device} We consider a dual-gated geometry (Fig.~\ref{fig:figone}b), in which the gate electron densities $n_t,n_b$ can be controlled, allowing for independent control of total sample electron density
$n=-n_t-n_b$ and displacement field $D=e(n_b-n_t)/(2\epsilon_0)$.
Within the standard theoretical approach, the effect of the displacement field is modeled as giving
a constant layer-potential drop 
$U_{l+1}-U_l= -\Delta U = - d_leD/(\epsilon_\perp)$,
with $d_l$ the interlayer distance and $\epsilon_\perp$ the out-of-plane dielectric constant. This model corresponds to the potential drop in the absence of the sample and thus ignores the self-consistent screening physics discussed in the Introduction. Therefore, we instead use Gauss' law directly
\footnote{
We note that our description here centers on a capacitor-plate model for clarity, but it could also be obtained by considering the standard density-
density interaction and incorporating the layer dependence of
the Coulomb interaction. In that viewpoint, the layer-potential terms arise from the $q=0$ Hartree contribution of charges localized in different layers. Eq. \eqref{eq:potdifference} makes clear that the layer potential is generically non-linear in $l$, rather than a constant-gradient ramp, which is essential for the near-surface drop seen in Fig.~\ref{fig:figone}e-g. See Refs. \cite{Kolar2026PRB_SingleGateTracking,lewandowskiKolarElectrostaticFate$N$layer2023,2025arXiv250507981T} for further discussion.}, which gives the out-of-plane field between layers $l$ and $l+1$,
\begin{equation}
E^{\perp}_{l,l+1}=\frac{e}{\epsilon_\perp\epsilon_0}\sum_{j=0}^{l} n_j,
\end{equation}
where $n_j$ is the net \textit{electron} number density on layer $j$, with $n_0\equiv n_b$ and $n_{N+1}\equiv n_t$ so the same expression applies to the gate--sample regions.
The associated electrostatic energy density is
\begin{equation}
\label{eq:potenergy}
\mathcal{E}_{PE}=\sum_{l=1}^{N-1}\epsilon_\perp\epsilon_0\, d_l\,\frac{\big(E^{\perp}_{l,l+1}\big)^2}{2},
\end{equation}
and the potential differences experienced by the electrons [entering Eq.~\eqref{eq:hamrhombo}] are 
\begin{equation}
\label{eq:potdifference}
U_{l+1}-U_l=-e^2 d_l\,\frac{\sum_{j=0}^{l} n_j}{\epsilon_\perp\epsilon_0}.
\end{equation}
Given particular gate densities $(n_b,n_t)$, we obtain self-consistent band structures by iterating Eqs.~\eqref{eq:hamrhombo} and \eqref{eq:potdifference} to convergence (see Supplement~\cite{supplement}); unless noted, we use $\epsilon_\perp=3$ \cite{Slizovskiy2021}.

The inclusion of electrostatic energy cost (Eq.~\eqref{eq:potenergy}) in theoretical modeling is an essential component in determining the phase diagram of thick stacks. It drives the complete screening of the top (bottom) gate effect on the bottom (top) surface for $N\to \infty$,
because gating works by generating an electric field, and allowing the gate field to penetrate the thick stack would incur an electrostatic energy penalty 
$\mathcal{E}_{PE}$ that grows with the number of layers $\mathcal{E}_{PE}\propto N-1$.
However, accounting for this cost is also important for the physics close to the surfaces. 
In particular, Eq.~\eqref{eq:potenergy} dictates that the bottom (top) layer polarized states at small $|\mathbf k|$ respond more to tuning $n_b$ ($n_t$) than vertically delocalized states, which are effectively screened by the polarized pocket,
as discussed in Ref.~\cite{Kolar2026PRB_SingleGateTracking}.

{\bf Electrostatic recovery of flat bands for $N\to \infty$}
  How do self-consistent layer potentials change the non-interacting dispersion? For the layer distributions from Eq.~\eqref{eq:wlpuremodel}, the interaction-induced dispersion is generally 
 \begin{equation}
 \label{eq:intinduceddisp}
 E^{\text{int.}}_{|\mathbf k| = \frac{t_1}{v_F}\sqrt{x}} = \sum_l W_l(x) U_l =
 U_1+ \sum_{l=1}^{\infty} (U_{l+1}-U_l) x^{l}.
 \end{equation}
The total dispersion of the surface band, including the layer potentials, can then be expressed as
a sum of the zero potential result plus a potential dependent shift:
\begin{eqnarray}
\label{eq:fullpotdispersion}
E_{\mathbf k}& \approx& E_{\mathbf k}|_{U_l=0}  + E^{\text{int.}}_{\mathbf k} \nonumber \\
& = & 2t_1 \frac{v_4}{v_F}x + \sum_{l=1}^{\infty} (U_{l+1}-U_l) x^{l},
\end{eqnarray}
where $x<1$ defines the surface band.
The above estimate highlights the condition for optimal band flattening: a large near-surface drop $U_2-U_1=  -2t_1(v_4/v_F)\approx -\SI{40}{meV}$ with vanishing deeper-layer drops ($U_{l+1}-U_l\approx 0$ for $l>2$).

We first consider charge neutrality (CN), where the gates are uncharged.
Our analytical picture comes from two controlled limits. \emph{Strong-coupling, isotropic limit.} In this limit ($1/\epsilon_\perp\!\to\!\infty$), a simplified surface-band model (using the dispersion in Eq.~\eqref{eq:fullpotdispersion} for $x<1$) admits a self-consistent solution in which all surface-band states are half-filled and the mean-field dispersion becomes exactly flat (i.e., $\mathbf{k}$ independent).
Within Eq.~\eqref{eq:fullpotdispersion}, this occurs when the only nonzero layer-potential drop is
$U_2-U_1=-2t_1(v_4/v_F)$, while all deeper drops vanish ($U_{l+1}-U_l\approx0$ for $l>2$).
A minor technical subtlety is that realizing this condition requires a slight departure from perfectly uniform half-filling; we explain this in the Supplement~\cite{supplement}.
Including trigonal warping does not change the basic mechanism, but it prevents \emph{perfect} cancellation: layer potentials cannot remove the angular anisotropy, so an irremovable minimal bandwidth remains, set by the warping. \emph{Weak-coupling limit.}
In the opposite limit of weak interactions (large $\epsilon_\perp$), CN fills the surface states up to $x=1/2$ on each surface.
This produces excess charge concentrated on the outermost layers and therefore a surface-localized electrostatic profile.
Near the bottom surface one finds analytically (Supplement~\cite{supplement})
\begin{equation}
U_{l+1}-U_l=\frac{e^2 d_l n_c}{2\epsilon_\perp\epsilon_0}\frac{1}{l+1}
\left[\left(\frac{1}{2}\right)^l-1\right],
\end{equation}
which already counteracts the quadratic $v_4$-induced dispersion and begins to flatten the band.
As the interaction strength is increased, self-consistent electrostatic relaxation makes the potential decay away from the surface more rapidly (from power-law to exponential; Fig.~\ref{fig:figone}f, bottom), and the bandwidth decreases monotonically toward the maximally flat strong-coupling limit (up to the trigonal-warping-imposed minimum).
The above monotonic trend can be further explained by two general constraints discussed in the Supplement~\cite{supplement}:
(i) if the $\mathbf{k}=0$ state is occupied, the dispersion at $\mathbf{k}=0$ must have positive curvature, and
(ii) at CN the $\mathbf{k}=0$ state must be occupied. These analytical expectations are supported by fully self-consistent calculations for $N=50$ RNG.
Already at $\epsilon_\perp=3$ the CN band in Fig.~\ref{fig:figone}f shows substantial flattening relative to $U_l=0$.
Moreover, increasing the interaction strength produces systematically flatter dispersions in our numerical simulations (see the Supplement~\cite{supplement}).

\begin{figure*}
    \centering
    \includegraphics[width=2\columnwidth]{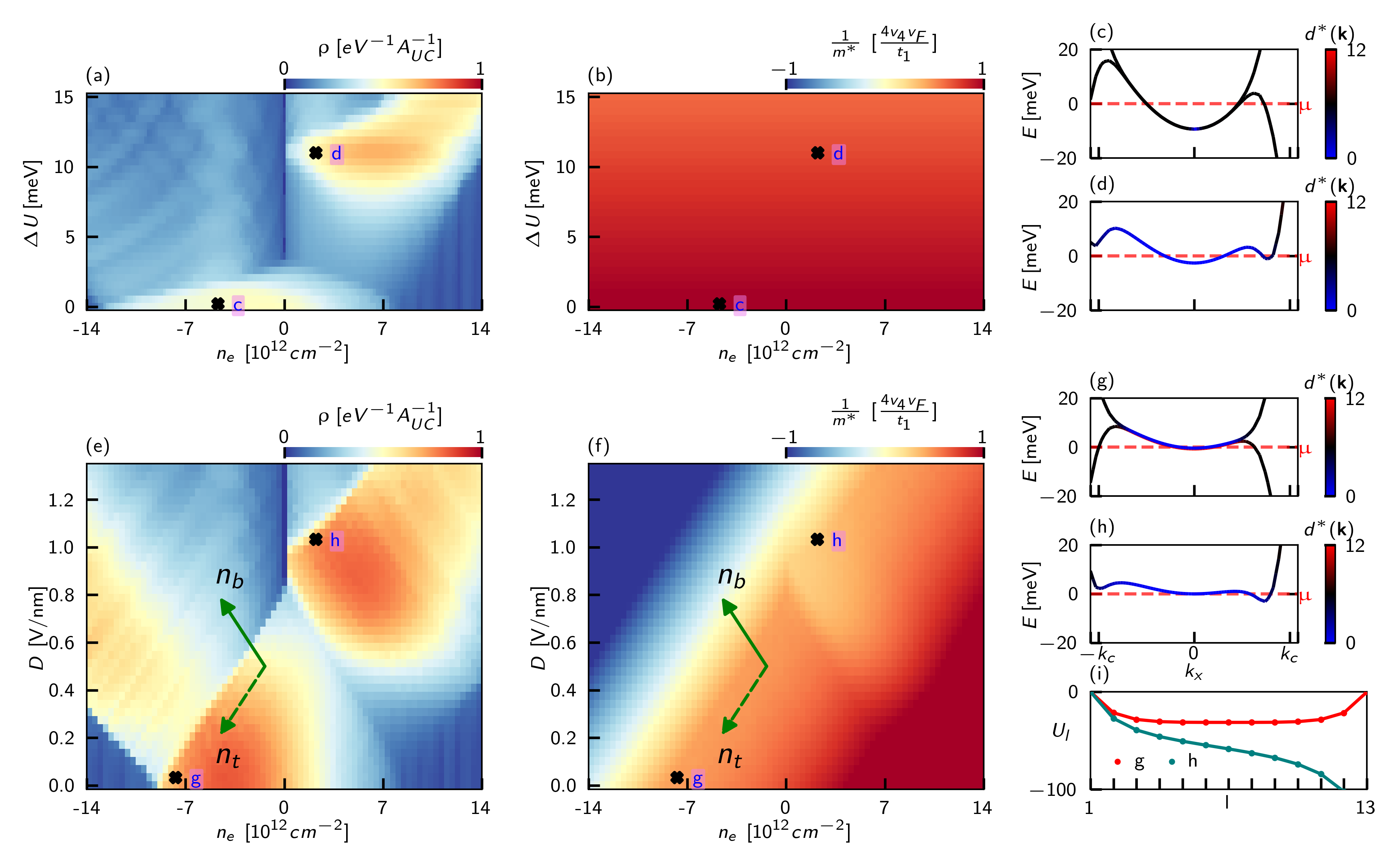}
    \caption{(a) $n-\Delta U$ map of the density of states (DOS) calculated for a model where the potential difference between adjacent layers is assumed fixed, $U_{l+1}-U_l=-\Delta U$.
    (b) Effective mass, shown for the same fixed potential-difference model as in (a), in units of the mass defined in Eq.~\eqref{eq:zeropotdispersion}.
    (c) Band structure from the fixed potential-difference model described in (a), shown here at zero potential. (d) Band structure from the same model at a large potential, focusing on the region with high DOS.
(e) $n-D$ map of the DOS for the full self-consistent electrostatic model, using potentials from Eq.~\eqref{eq:potdifference}. (f) Effective mass for the same self-consistent model as in (e).
    (g) Band structure in the coexistence region
    (h) Band structure from the self-consistent model, shown at high $D$.
    (i) Potential ($U_l$) profile calculated for the band structures in (g) and (h) under the self-consistent model.
}
    \label{fig:figthree}
\end{figure*}

Next, we explain how gating modifies the electrostatically induced flattening of the surface band. 
Hole doping by the proximal gate preferentially depletes the layer-polarized pocket near $\mathbf{k}=0$, which screens the gate field most efficiently and thus avoids a large Coulomb cost. Because this pocket is nearly bottom-layer polarized, its depletion primarily increases the potential drop between the first two layers (compare bottom panels of Figs.~\ref{fig:figone}e,f), moving the system toward the configuration that optimizes flattening, while the pocket must retain positive curvature 
at $\mathbf{k}=0$ so long as it is occupied [point (i) above]. This continues until the $\mathbf{k}=0$ pocket empties, defining the lower band edge $n_-$ in Fig.~\ref{fig:figone}a. For $n<n_-$ the surface band rapidly moves away from the Fermi level as the remaining finite-$\mathbf{k}$ states are more layer-delocalized and screen the gate electric field poorly. Notably, the strongest flattening occurs near this depletion point.
We confirm this trend numerically, showing the self-consistent, strongly flattened, dispersion near the depletion point in
Fig.~\ref{fig:figone}e.
The opposite trend appears near full filling ($n\approx 2n_+$): the potential becomes inverted-$U$ shaped and the band becomes more dispersive (Fig.~\ref{fig:figone}g). Even there, however, the overall dispersion is still set primarily by electrostatics rather than the bare $v_4$ term (cf. Fig.~\ref{fig:figone}b).

{\bf Finite $N$ effects} 
We now turn to finite-thickness corrections away from the $N\!\to\!\infty$ limit, focusing on the experimentally relevant case $N=13$ \cite{Guo2025arXiv_ThickRhombohedral_SC}, shown in Fig.~\ref{fig:figthree}. At this thickness, the surface bands remain well-defined, but the two surfaces are no longer electrostatically independent. 
The non-interacting band structure at $U_l=0$ (Fig.~\ref{fig:figthree}c) shows well-defined surface bands,
that, however merge into the bulk at smaller momenta than in the large-$N$ limit. Correspondingly, the self-consistent DOS map in the $n$--$D$ plane (Fig.~\ref{fig:figthree}e) still displays the familiar coexistence ``diamond,'' but with clear departures from the ideal large-$N$ ``stripe'' picture of Fig.~\ref{fig:figone}a: the bottom-surface stripe becomes discontinuous across $n=0$, and its upper boundary is no longer perfectly gate-tracking. Physically, this reflects finite-$N$ cross-talk: even the distant gate and the opposite surface contribute appreciably to the electrostatics of a given surface band.

This cross-talk is most apparent when comparing band structures in the coexistence region—where both surfaces meet the Fermi level—with those in the high-$D$ regime, where one surface is depleted. In the coexistence region (Fig.~\ref{fig:figthree}g), the self-consistent potentials are relatively weak and spread across many layers (red curve in Fig.~\ref{fig:figthree}i). As a result, finite-$\mathbf{k}$ states are only partly pulled toward the Fermi level: the band is flatter than at $U_l=0$ (compare to Fig.~\ref{fig:figthree}c), but not completely flat. By contrast, at high $D$, the top surface is depleted, resulting in a sharper potential drop near the surface (teal curve in Fig.~\ref{fig:figthree}i) as the electric field penetrates the stack. This strong, localized drop lowers finite-$\mathbf{k}$ states more and produces a much flatter bottom-surface band (Fig.~\ref{fig:figthree}h). This is essentially the flattening mechanism known for small $N$.

It is instructive to compare these fully self-consistent results to the commonly used linear-potential approximation,
$U_{l+1}-U_l=-\Delta U=-eDd_l/(\epsilon_0\epsilon_\perp)$.
While the linear model captures the broad $n$--$\Delta U$ features (Fig.~\ref{fig:figthree}a) and can flatten parts of the band at large $\Delta U$ (Fig.~\ref{fig:figthree}d), it stands in contrast to the self-consistent calculation by missing the strong renormalization of curvature near $\mathbf{k}=0$. Specifically, the inverse effective mass changes only minimally in the linear model (Fig.~\ref{fig:figthree}b) compared to the larger variations in the self-consistent result (Fig.~\ref{fig:figthree}f), and band dispersions remain generally more curved in the linear model across relevant parameter ranges (see Fig.~\ref{fig:figthree}c,d,g,h). For $N=13$, this means a linear $U_l$ profile (i) does not capture the significant $\mathbf{k}=0$ flattening tied to a sharp near-surface drop, and (ii) misses the extra flattening near $n<0$, $D\approx 0$ that arises from gate charges $n_{b,t}\approx -n_-$. 

Finally, we note that the $v_4$-induced quadratic dispersion of the $\mathbf{k}=0$ pocket produces a visible particle--hole asymmetry within the coexistence diamond, seen in Fig.~\ref{fig:figthree}e. The left boundary near $n_-$ corresponds to a sharp onset of high DOS when a layer-polarized $\mathbf{k}=0$ pocket enters the Fermi level, whereas the right boundary near $n_+$ is smoother because it is controlled by more layer-delocalized finite $\mathbf k$ states exiting the Fermi level (Fig.~\ref{fig:figthree}e,f). In a particle--hole-symmetric model (without an intrinsic quadratic edge), these two boundaries would be approximately mirror-symmetric, and the DOS would vary more smoothly rather than showing a sharp step associated with parabolic dispersion.

{\bf Discussion} 
Our central result is that nonlinear, self-consistent electrostatics provides a robust route to recovering surface-band flatness in rhombohedral $N$-layer graphene, even in the absence of an applied displacement field. In the material-accurate model, remote (PH-breaking) hoppings ($v_4$) generate a quadratic dispersion at $U_l=0$ (Fig.~\ref{fig:figone}b), but electrostatic forces generate a near-surface electrostatic potential whose curvature opposes the $v_4$-induced dispersion contribution (Fig.~\ref{fig:figone}e,f). Crucially, this mechanism is \emph{self-stabilizing}: screening concentrates the potential drop toward the outermost layers and prevents runaway ``over-bending,'' consequently strengthening the flattening as interactions increase.

Importantly, our theory predicts that a large DOS (density of states) can already occur at $D=0$ on the hole-doped side, where electrostatic potentials strongly pull finite-$k$ states (states with nonzero momentum) toward the Fermi level (Fig.~\ref{fig:figone}c,d). This hints at a microscopic explanation for the low-$D$ interaction-driven regime observed in recent experiments \cite{Guo2025arXiv_ThickRhombohedral_SC}.

Beyond the present experimental phenomenology, the large-$N$ limit points to a distinct material design opportunity. Increasing $N$ increases the number of available surface states and hence it also enhances susceptibility to interaction-driven instabilities at small applied fields. In this sense, electrostatic flattening at $D\approx 0$ offers a complementary route to strong correlations without relying on large displacement fields. This suggests that thicker flakes should generally support broader regions of symmetry breaking, as well as potentially higher energy scales for ordered phases once the surface bandwidth is sufficiently suppressed. 

As the relevant low-energy states are extremely layer polarized near $\mathbf k=0$, any correlated order that couples to the layer degree of freedom should be especially favorable on this platform: layer selectivity favors layer-selective ordering by reducing the number of active degrees of freedom and enabling strong coupling to gate-tunable electrostatics. This perspective naturally connects to the growing experimental evidence (e.g. Refs. \cite{mishchenkoShiElectronicPhaseSeparation2020,falkoSlizovskiyFilmsRhombohedralGraphite2019,PhysRevB.81.161403,PhysRevB.97.245421,Zhou2021,Zhou2021-b,Liu2023,Han2025,Auerbach2025,Han2023,Guo2025arXiv_ThickRhombohedral_SC,Kumar2025arXiv_DualSurface_SC,Deng2026arXiv_Hexalayer_Semimetal,2025arXiv250405129Q,Zhou2024}) for electrically tunable magnetic order in rhombohedral multilayers, including ferroelectric/orbital-magnetic phenomena and superconductivity emerging from a correlated semimetal background, and motivates viewing layer-selective orders as natural competitors or parents to the superconducting states seen experimentally.

Our work highlights a useful separation of roles: electrostatics primarily reshapes the dispersion by generating a nonlinear near-surface potential profile, while leaving the underlying surface-state wavefunctions —and their quantum geometry
\cite{f43d-414n,kwanBernevigBerryTrashcanModel2025,pgqh-wm5v} —as central once the band is sufficiently flat. As a result, electrostatic flattening at low $D$ opens a promising pathway for systematic control of electronic phases sensitive to geometry, such as fractionalized and topological orders proposed for rhombohedral surface states (e.g. Refs. \cite{PhysRevB.109.205121,PhysRevB.109.205121,PhysRevB.109.205122,PhysRevLett.133.206502,PhysRevB.112.075109,PhysRevLett.128.176404}). Our work thus motivates future experiments combining layer-resolved probes with controlled inter-surface coupling, to explore the crossover from two nearly independent surfaces at $D=0$ to the single-surface flat-band regime at large $|D|$.

\section{Acknowledgments}
Numerical calculations are done using the High Performance Compute Cluster of the Research Computing Center (RCC) at Florida State University. K.K. was supported by a grant from the Simons Foundation (SFI-MPS-NFS-00006741-12, P.T.) in the Simons Collaboration on New Frontiers in Superconductivity.   A.F.Y. was primarily supported by the Department of Energy under Award DE-SC0020043 to A.F.Y., and by the Gordon and Betty Moore Foundation, grant DOI 10.37807/GBMF13801. C.L. was supported by start-up funds from Florida State University and the National High Magnetic Field Laboratory. 
The National High Magnetic Field Laboratory is supported by the National Science Foundation through NSF/DMR-2128556 and the State of Florida.

\IfFileExists{klibrary.bib}{\bibliography{klibrary.bib,additional.bib}}{\IfFileExists{../klibrary.bib}{\bibliography{../klibrary.bib,../additional.bib}}{}}

\begin{widetext}
\pagebreak

\onecolumngrid

\setcounter{secnumdepth}{2}
\begin{center}
    \large \bfseries Supplementary material
\end{center}

\renewcommand{\theequation}{S\arabic{equation}}
\setcounter{equation}{0}

\renewcommand{\thefigure}{S\arabic{figure}}
\setcounter{figure}{0}

\tableofcontents
\section{Additional details}
\subsection{Energy shift of a given state for a given potential profile}

We show that the interaction-induced energy shift can be written entirely in terms of
layer-to-layer potential drops. In the particle--hole (PH) model, the normalized layer
weight of a surface state with in-plane momentum $\mathbf{k}$ is
\begin{equation}
\label{eq:wlpuremodelsuppl}
W_l(\mathbf{k})=(1-x)x^{l-1},\qquad
x=\left(\frac{v_F|\mathbf{k}|}{t_1}\right)^2,
\end{equation}
so that the interaction shift is $E^{\text{int}}(\mathbf{k})=\sum_l U_l\,W_l(\mathbf{k})$. Then
\begin{align}
E^{\text{int}}(x)
&=\sum_{l\ge1} U_l\,W_l(x)
 =\sum_{l\ge1} U_l\,(1-x)x^{l-1} \nonumber\\
&=\sum_{l\ge1} U_l\,(x^{l-1}-x^l)
 =\sum_{l\ge1}U_l x^{l-1}-\sum_{l\ge1}U_l x^l \nonumber\\
&=U_1+\sum_{l\ge1}U_{l+1}x^l-\sum_{l\ge1}U_l x^l
 =U_1+\sum_{l\ge1}(U_{l+1}-U_l)\,x^l .
\end{align}

\subsection{Failure of a linear potential to flatten the bands at $N=50$}

\begin{figure*}[t]
    \centering
    \includegraphics[width=0.9\textwidth]{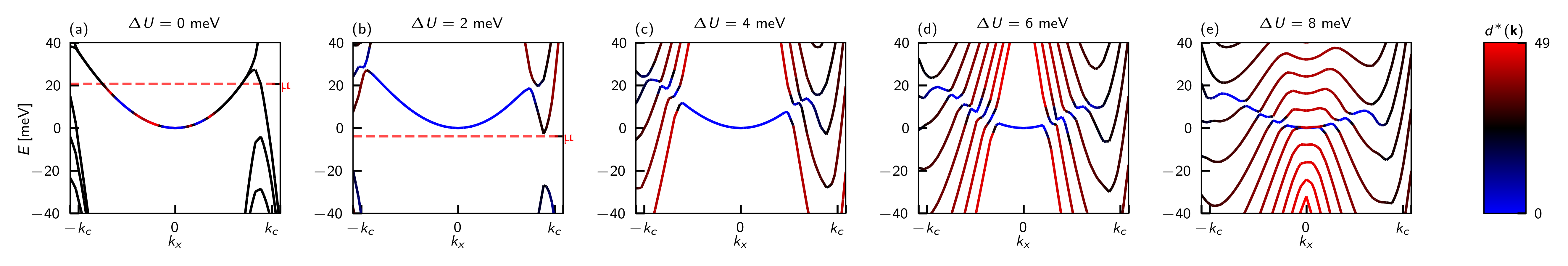}
    \caption{Band structures for $N=50$ computed using a linear layer potential $U_l=-l\,\Delta U$.}
    \label{fig:suppllinearpotfifty}
\end{figure*}

To assess the commonly used linear-potential approximation in the large-$N$ limit,
we evaluate Eq.~\eqref{eq:fullpotdispersion} for a linear profile $U_l=-l\,\Delta U$.
This gives
\begin{align}
\label{eq:dispersionlinear}
E_{\mathbf{k}}
&= 2t_1\frac{v_4}{v_F}\,x \;-\; \Delta U\,\frac{x}{1-x} \nonumber\\
&= x\left[\,2t_1\frac{v_4}{v_F}\;-\;\frac{\Delta U}{1-x}\right],
\end{align}
where $x=(v_F|\mathbf{k}|/t_1)^2$ parameterizes the surface-band states ($0\le x<1$).
A key feature is the pole at $x\to 1$, which signals strong distortion and mixing near the edge of the surface-band domain.

Figure~\ref{fig:suppllinearpotfifty} shows the resulting $N=50$ band structures for $\Delta U\in[0,\SI{8}{meV}]$.
Already at $\Delta U=\SI{4}{meV}$ the linear profile induces substantial band mixing and the surface band is no longer well isolated.
Moreover, at charge neutrality the surface states are pushed far from the Fermi level in this model
(see Fig.~\ref{fig:suppllinearpotfifty}c--e, where the Fermi level lies outside the plotted window).

\subsection{Analytical expression for $U$-shaped potential at charge neutrality without self-consistency}
\label{sec:supplanalyticalushape}

We consider a \emph{frozen-charge} construction: we fill the noninteracting ($U_l=0$) surface band by occupying only states with $x<y$ (with $x$ defined in Eq.~\eqref{eq:wlpuremodelsuppl}), and we do \emph{not} allow the electrons to relax self-consistently. This corresponds to filling the $U_l=0$ bands in Fig.~\ref{fig:figone}d. Our main interest is charge neutrality, which in this parametrization corresponds to half-filling, $y=\tfrac12$.

To properly account for the remote-band background, we use that there is no excess charge on any layer when each momentum state in the surface band is half-filled.
This is the situation in the PH symmetric model, which has no charge inhomogeneity by symmetry.
For the realistic model considered here, we consider that the wavefunctions are modified only slightly compared
to the PH model, so that the same applies.
We therefore subtract the half-filled layer density
\begin{align}
n_l^{0}
&=\frac{n_c}{2}\int_{0}^{1}\!dx\,W_l(x)
=\frac{n_c}{2}\left(\frac{1}{l}-\frac{1}{l+1}\right),
\end{align}
where $W_l(x)$ is given in Eq.~\eqref{eq:wlpuremodelsuppl}. With this convention, the layer density obtained by filling only states with $x<y$ is
\begin{align}
n_l^{x<y}
&=n_c\int_{0}^{y}\!dx\,W_l(x)-n_l^{0} \nonumber\\
&=\frac{n_c}{2}\int_{0}^{y}\!dx\,W_l(x)-\frac{n_c}{2}\int_{y}^{1}\!dx\,W_l(x) \nonumber\\
&=n_c\left(\frac{y^{l}}{l}-\frac{y^{l+1}}{l+1}\right)
-\frac{n_c}{2}\left(\frac{1}{l}-\frac{1}{l+1}\right).
\end{align}
The corresponding bottom-gate charge density is
\begin{equation}
n_b=-y\,n_c+\frac{n_c}{2},
\end{equation}
so that $y=\tfrac12$ indeed gives $n_b=0$ (uncharged gates at charge neutrality).

Using Eq.~\eqref{eq:potdifference}, the resulting (non-self-consistent) potential drop is
\begin{equation}
\label{eq:potdiffnonselfconsistent}
U_{l+1}-U_l=
\frac{e^2 d_l n_c}{\epsilon_\perp\epsilon_0}
\left[\frac{y^{l+1}}{l+1}-\frac{1}{2}\frac{1}{l+1}\right].
\end{equation}
As shown in Fig.~\ref{fig:supplfignonselfconsist}a, Eq.~\eqref{eq:potdiffnonselfconsistent} captures the \emph{shape} of the induced near-surface profile at charge neutrality (in particular its $U$-like form), consistent with the trend seen in the full self-consistent calculations.

However, this frozen-charge estimate substantially overestimates the field deep in the stack: for large $l$ it decays only as $\sim 1/(l+1)$ (a power-law tail), whereas the fully self-consistent solution exhibits an exponential decay of the electric field into the bulk. This contrast is illustrated in Fig.~\ref{fig:supplfignonselfconsist}b, where we compare the non-self-consistent profile to the fully self-consistent profile at two representative points in the phase diagram.

\begin{figure*}[t]
    \centering
    \includegraphics[width=0.45\textwidth]{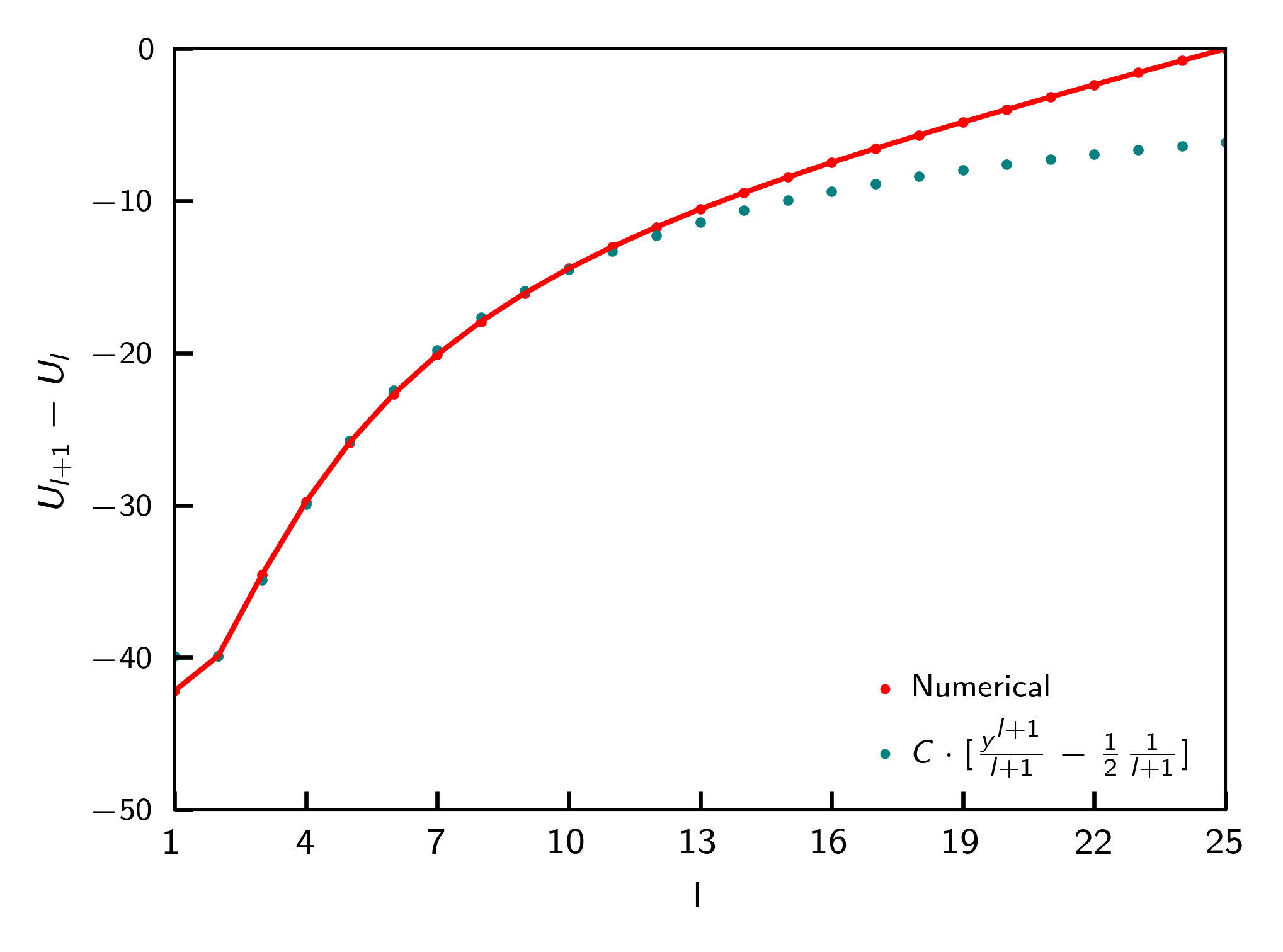}
    \hfill
    \includegraphics[width=0.45\textwidth]{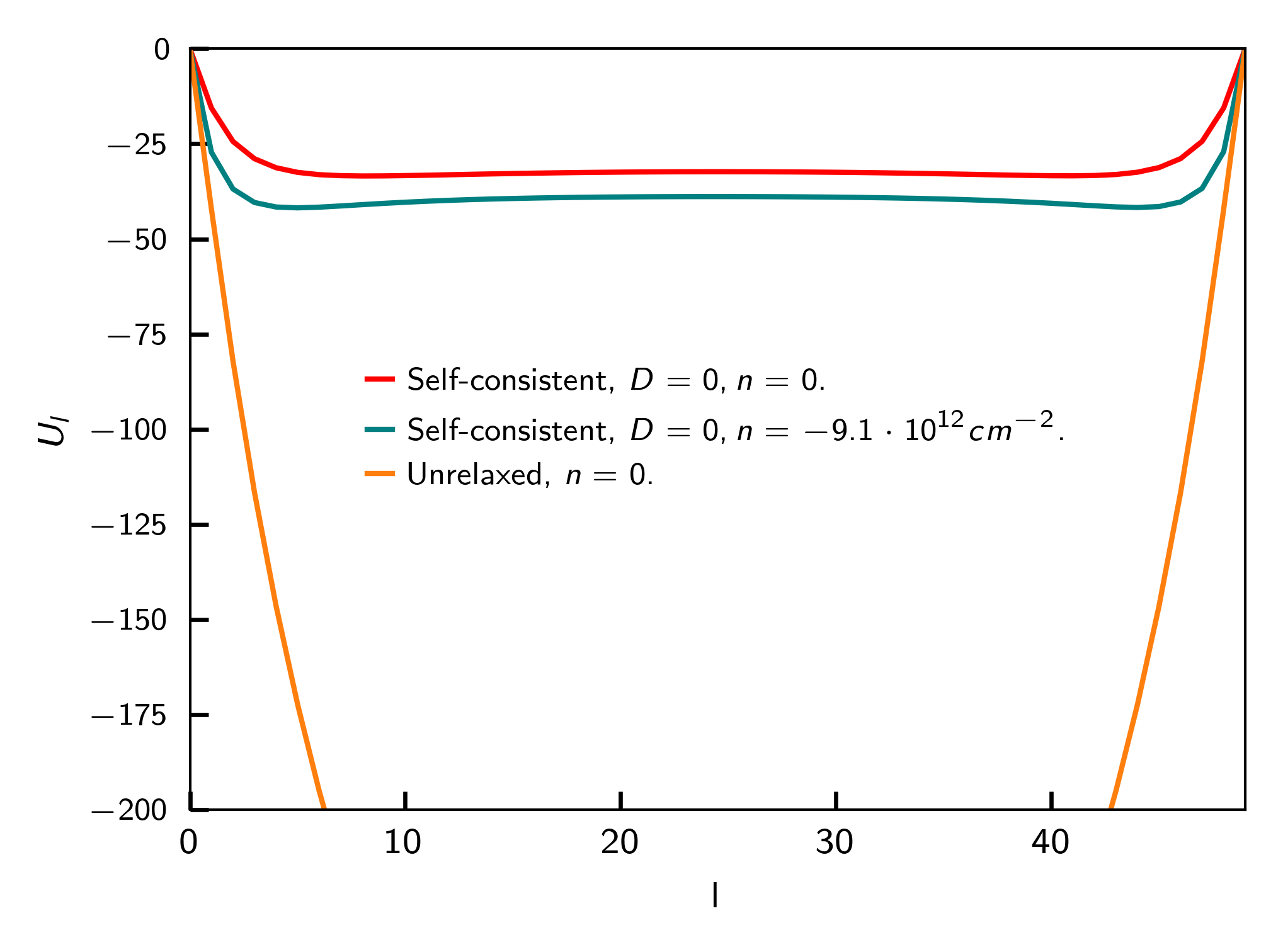}
    \caption{
    (a) Frozen-charge (non-self-consistent) potential drops $U_{l+1}-U_l$ at $n=0$ computed from the non-interacting ($U_l=0$) band filling (red), compared with the analytical form in Eq.~\eqref{eq:potdiffnonselfconsistent} (green), shown with a single overall scale factor.
    (b) Comparison of layer potentials from the frozen-charge construction and the fully self-consistent calculation for $N=50$.
    }
    \label{fig:supplfignonselfconsist}
\end{figure*}

\subsection{Exponentially decaying electric fields with self-consistency}

Self-consistency qualitatively changes the long-distance behavior of the electrostatic profile. 
In particular, it removes the slow power-law tail of the frozen-charge estimate and yields an
approximately exponential decay of the layer-to-layer potential drops into the bulk,
\begin{equation}
U_{l+1}-U_l \simeq C\,y^{\,l}\qquad (0<y<1),
\end{equation}
with $C$ a prefactor and $y$ the decay factor. Figure~\ref{fig:supplfigselfconsistfits} shows representative
self-consistent profiles together with best-fit exponentials.

\begin{figure*}[t]
    \centering
    \includegraphics[width=0.45\textwidth]{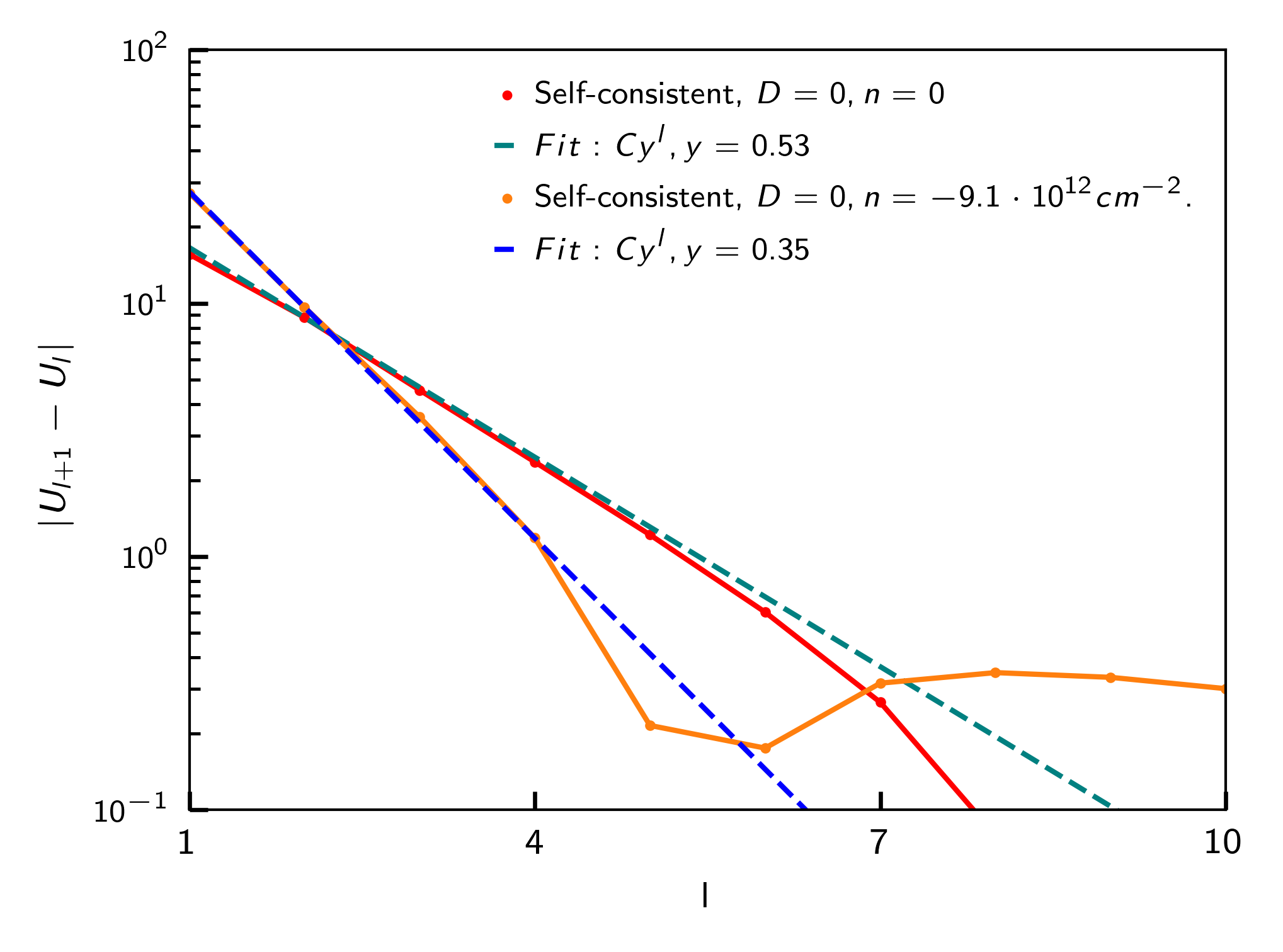}
    \caption{
    Self-consistent potential drops $U_{l+1}-U_l$ for $N=50$ at $n=0$ and $n=\SI{-9.1e12}{cm^{-2}}$ (solid),
    compared to exponential fits $C\,y^{\,l}$ (dashed).
    }
    \label{fig:supplfigselfconsistfits}
\end{figure*}

\subsection{Analytical arguments for stabilization of the flattening mechanism}

We provide two analytical observations that help explain why the self-consistent electrostatics
tends to \emph{flatten} the surface band at large $N$.
Throughout we assume, for simplicity, that the layer-to-layer potential drops are sign-definite
away from the bottom surface, i.e.\ $U_{l+1}-U_l$ has the same sign for all $l$ (a monotonic profile).

\paragraph*{(i) If $\mathbf{k}=0$ is occupied, its curvature must be positive (except at full filling).}
We argue that, away from full filling, the occupied band must have \emph{positive} curvature at $\mathbf{k}=0$
whenever the $\mathbf{k}=0$ state is occupied.
To see this, write the surface-band dispersion (Eq.~\eqref{eq:fullpotdispersion}) in the form
\begin{align}
\label{eq:fullpotdispersionsupplone}
E(\mathbf{k})
&\approx x\Bigg[
2t_1\frac{v_4}{v_F} + (U_2-U_1)
+ \sum_{l=1}^{\infty}(U_{l+1}-U_l)\,x^{l-1}
\Bigg],
\end{align}
with $x=(v_F|\mathbf{k}|/t_1)^2$.
The sign of the bracket at small $x$ controls the curvature at $\mathbf{k}=0$; in particular,
\begin{equation}
2t_1\frac{v_4}{v_F} + (U_2-U_1) < 0
\end{equation}
corresponds to negative curvature at $\mathbf{k}=0$.
If, moreover, the bracket in Eq.~\eqref{eq:fullpotdispersionsupplone} remains negative for all $0<x<1$,
then $x=0$ is a \emph{global maximum} of the surface band (since $E(0)=0$ and $E(x)<0$ for $x>0$),
so the $\mathbf{k}=0$ state would be empty at $T=0$ for any filling short of full filling.
For a monotonic $U$-profile with $U_{l+1}-U_l<0$, the sum in Eq.~\eqref{eq:fullpotdispersionsupplone}
can only decrease the bracket as $x$ increases, making this scenario natural.
More generally, the same conclusion holds for non-monotonic profiles whenever the bracket remains negative over $0<x<1$.
Therefore, if $\mathbf{k}=0$ is occupied at a filling other than full filling, the dispersion must have positive curvature at $\mathbf{k}=0$.

\paragraph*{(ii) At charge neutrality, $\mathbf{k}=0$ must be occupied.}
We next argue that at charge neutrality (CN) the $\mathbf{k}=0$ state cannot be self-consistently unoccupied.
Using Eq.~\eqref{eq:fullpotdispersion}, we rewrite the dispersion as
\begin{align}
\label{eq:fullpotdispersionsuppl}
E(\mathbf{k})
&\approx x\Bigg[
2t_1\frac{v_4}{v_F}
+ \sum_{l=1}^{\infty}(U_{l+1}-U_l)\,x^{l-1}
\Bigg],
\end{align}
so that $E(0)=0$ and occupied states at $T=0$ must satisfy $E(\mathbf{k})<0$.
For a monotonic profile there are two cases:

\emph{(a) $U_{l+1}-U_l>0$:}
then the bracket in Eq.~\eqref{eq:fullpotdispersionsuppl} is positive at small $x$, so states near $\mathbf{k}=0$
are among the lowest-energy states and $\mathbf{k}=0$ is trivially occupied.

\emph{(b) $U_{l+1}-U_l<0$ (a $U$-shaped profile):}
then the bracket in Eq.~\eqref{eq:fullpotdispersionsuppl} is either negative for all $x$ (negative effective mass)
or becomes negative only beyond some $x$.
In either situation, forcing $\mathbf{k}=0$ to be empty implies that the lowest-energy occupied states lie at finite momentum,
and the resulting filling necessarily weights larger $x$ more strongly (schematically, occupations pushed toward $x\gtrsim 1/2$).
However, occupying predominantly larger-$x$ states produces an \emph{inverted} potential profile (opposite sign of $U_{l+1}-U_l$),
since these states are more layer-delocalized.
This contradicts the assumed monotonic $U$-profile with $U_{l+1}-U_l<0$.
Hence, a self-consistent CN solution cannot have $\mathbf{k}=0$ unoccupied.

Together, (i) and (ii) constrain the self-consistent dispersion near CN: $\mathbf{k}=0$ must be occupied, and therefore
its curvature must remain positive (except at full filling). These constraints underpin the stability and monotonicity of the
electrostatically induced flattening discussed in the main text.

\subsection{Analytical flat band in the infinite-interaction limit}

A series of key insights to the flattening mechanism can be obtained from a toy model that retains only the surface band and uses the analytical dispersion, Eq.~\eqref{eq:fullpotdispersion}, for $0\le x<1$.
We use this model to show that in the $N\to\infty$ limit, the surface band approaches a perfectly flat mean-field dispersion in the strong out-of-plane interaction limit,
\begin{equation}
\frac{1}{\epsilon_\perp}\to\infty \qquad \Leftrightarrow \qquad \epsilon_\perp\to 0.
\end{equation}

In this limit, electrostatics strongly penalizes any layer charge imbalance, so the energetically preferred configuration at charge neutrality is to keep every layer (approximately) neutral. In the surface-band language this corresponds to making the occupation of \emph{each} momentum state $1/2$, which is exactly the situation at CN in the PH-symmetric model discussed in Sec.~\ref{sec:supplanalyticalushape}. However, such a momentum-independent occupation is only possible at $T=0$ if all momenta are degenerate, i.e.\ if the mean-field dispersion is exactly flat. Within Eq.~\eqref{eq:fullpotdispersion}, exact flatness ($\bm k$ independence) requires the near-surface drop to satisfy
\begin{equation}
U_2-U_1 = -2t_1\frac{v_4}{v_F}.
\end{equation}

At first sight, a \emph{finite} potential drop appears to contradict perfect charge homogeneity. The resolution is that for any \emph{finite} interaction strength there is a small but nonzero layer charge inhomogeneity, and the required inhomogeneity vanishes in the strong-coupling limit. Indeed, using Gauss' law for the near-surface drop,
\begin{equation}
U_2-U_1 = -\frac{e^2 d_l}{\epsilon_\perp\epsilon_0}\,n_1 = -2t_1\frac{v_4}{v_F},
\end{equation}
so that $n_1\propto \epsilon_\perp \to 0$ as $\epsilon_\perp\to 0$.

\paragraph*{Construction at exact charge neutrality.}
One way to generate the required drop at CN is to slightly bias the occupation toward $\mathbf{k}=0$ while depleting a nearby shell, keeping the \emph{total} filling fixed at $1/2$.
For example, take states with $0<x<\delta$ to be fully filled and states with $\delta<x<2\delta$ to be empty, with the remaining states half-filled. This produces an approximate near-surface drop
\begin{equation}
U_2-U_1 \simeq -\delta^2\,\frac{e^2 n_c d_l}{2\epsilon_0\epsilon_\perp}.
\end{equation}
Imposing $U_2-U_1=-2t_1(v_4/v_F)$ then yields
\begin{equation}
\delta \simeq \sqrt{4 t_1\frac{v_4}{v_F}\,\frac{\epsilon_0\epsilon_\perp}{e^2 n_c d_l}},
\end{equation}
which vanishes as $\delta\propto \sqrt{\epsilon_\perp}$ in the infinite-interaction limit $\epsilon_\perp\to 0$.

\paragraph*{Construction with infinitesimal doping.}
Alternatively, allow an infinitesimal gate-induced doping $-n_b=\delta\,n_c$ with $\delta\ll 1$ (and $\delta\to 0$ as $\epsilon_\perp\to 0$). In this case the added charge enters the $\mathbf{k}=0$ pocket and fully fills states up to $x=\delta$, while the remaining states stay half-filled. The resulting drop is
\begin{equation}
U_2-U_1 \simeq -\delta^2\,\frac{e^2 n_c d_l}{4\epsilon_0\epsilon_\perp},
\end{equation}
so self-consistency $U_2-U_1=-2t_1(v_4/v_F)$ gives
\begin{equation}
\delta \simeq \sqrt{8 t_1\frac{v_4}{v_F}\,\frac{\epsilon_0\epsilon_\perp}{e^2 n_c d_l}},
\end{equation}
again vanishing as $\delta\propto \sqrt{\epsilon_\perp}$ for $\epsilon_\perp\to 0$.

In both constructions, the key point is the same: in the strong-coupling limit the self-consistent electrostatics drives the band toward a flat dispersion ($\bm k$ independence) so that the surface-band momenta can remain (approximately) half-filled without generating layer charge imbalance.

Finally, trigonal warping prevents \emph{perfect} flattening using layer potentials alone: the angular anisotropy cannot be canceled everywhere in $\mathbf{k}$-space. Nevertheless, the strong-coupling tendency to reduce the bandwidth carries over, and the corresponding ``infinite-interaction'' limiting behavior is visible in the full multi-band self-consistent calculation by filling a state at $\mathbf k$, while keeping 
the state at $-\mathbf k$ empty, so that on average they are half-filled.
We show the self-consistent results in Fig.~\ref{fig:infiniteintfixedpoint}, where we show the band structures and DOS for increasing interaction ($\epsilon_\perp=3,1,0.2$),
showing a clear tendency towards increasing flatness.

\begin{figure*}[t]
    \centering
    \includegraphics[width=0.9\textwidth]{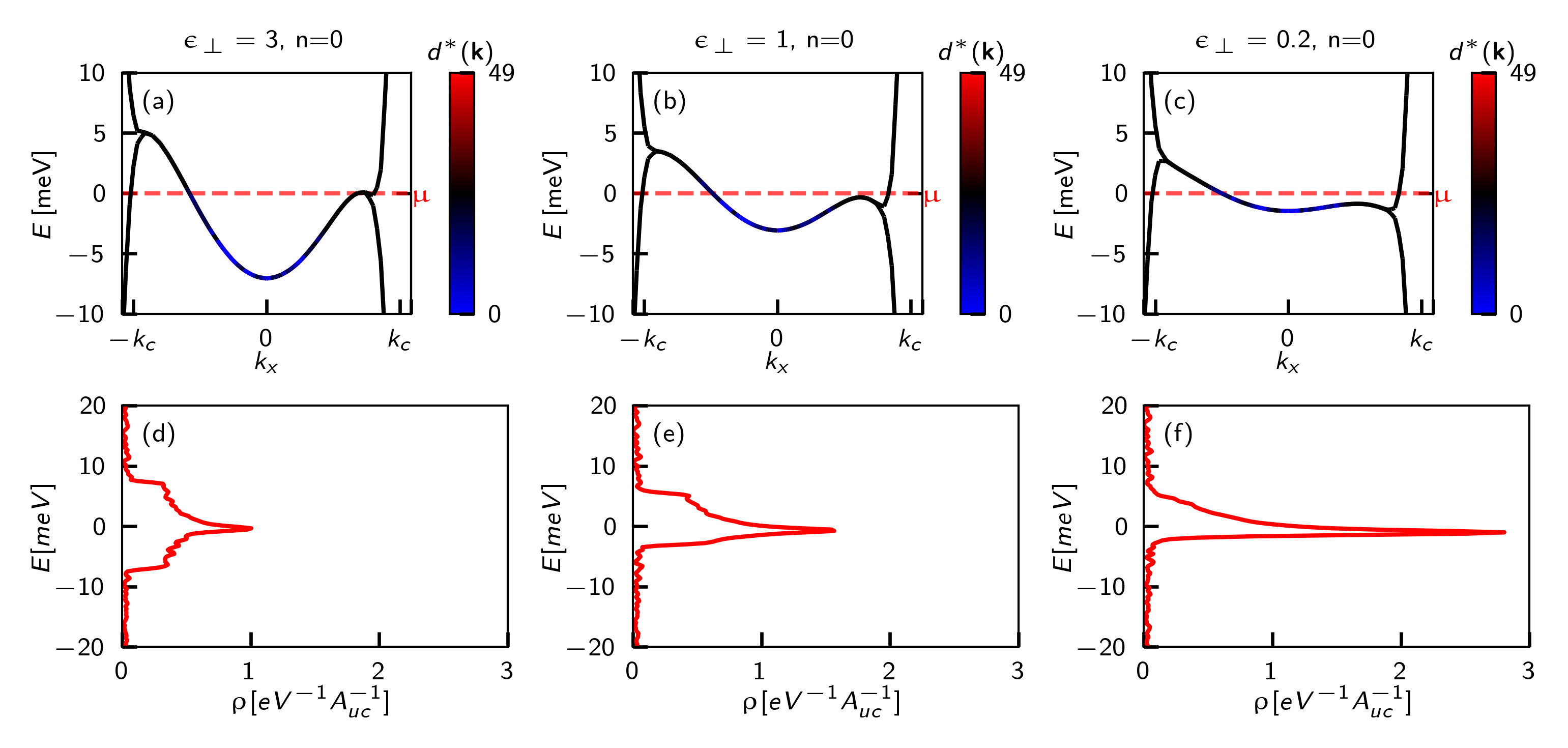}
    \caption{
    Charge neutrality ($n=0$, $D=0$) for $N=50$: interaction-enhanced flattening in the full self-consistent calculation, consistent with the strong-coupling trend toward a maximally flat surface band.
    }
    \label{fig:infiniteintfixedpoint}
\end{figure*}

\subsection{Electrostatics in the particle--hole-symmetric model}

To build intuition for the effect of gating and layer-dependent potentials, it is useful to study the particle--hole (PH) symmetric model. At $n=D=0$, each surface band is half-filled, the internal electric field vanishes, and the surface bands are perfectly flat (Fig.~\ref{fig:suppllphmodelbsfits}a).

Upon hole doping ($n<0$), carriers first occupy the most layer-polarized (small-$|\mathbf{k}|$) states because they screen the gate field most efficiently. By contrast, more layer-delocalized finite-$|\mathbf{k}|$ states screen less effectively and allow partial field penetration, which generates a $U$-shaped layer potential profile $U_l$. This $U$-shaped profile lowers the energies of finite-$\mathbf{k}$ states relative to $\mathbf{k}=0$, producing an interaction-induced negative-curvature bending of the surface band (Fig.~\ref{fig:suppllphmodelbsfits}b).

Neglecting band mixing, the interaction-induced shift of a surface state labeled by $x\propto|\mathbf{k}|^2$ (Eq.~\eqref{eq:wlpuremodelsuppl}) can be written as
\begin{equation}
\label{eq:supintinduceddisp}
E^{\mathrm{int}}(x)=\sum_l W_l(x)\,U_l
=U_1+\sum_{l=1}^{\infty}(U_{l+1}-U_l)\,x^{l},
\end{equation}
where $x=\left(\frac{v_F|\mathbf{k}|}{t_1}\right)^2$ and $W_l(x)$ is the normalized layer weight. In the PH-symmetric model the kinetic contribution vanishes, so the full dispersion is purely interaction-induced. As shown in Fig.~\ref{fig:suppllphmodelbsfits}, Eq.~\eqref{eq:supintinduceddisp} agrees well with the numerical dispersion.

As the hole density is decreased further, the self-consistent surface band ceases to remain flat near $n\simeq 2n_-$ (Fig.~\ref{fig:suppllphmodelbsfits}c). Notably, $|n_-|<n_c/2$, so the flat-band regime ends before the complete depletion of the noninteracting surface states at $n=-n_c$. By PH symmetry the upper boundary satisfies $2n_+=-2n_-$. As a result, in the $(n,D)$ plane each surface forms a gate-tracking stripe with $-n_b\in[n_-,n_+]$ (bottom surface) and $-n_t\in[n_-,n_+]$ (top surface), whose overlap produces the coexistence ``diamond'' schematized in Fig.~\ref{fig:figone}a. In this simple model, $n_\pm$ are set by the interaction-induced dispersion, emphasizing that the experimentally relevant flat-band filling range is generally reduced from $n_c$ and depends on interaction strength.

\begin{figure*}[t]
    \centering
    \includegraphics[width=0.9\textwidth]{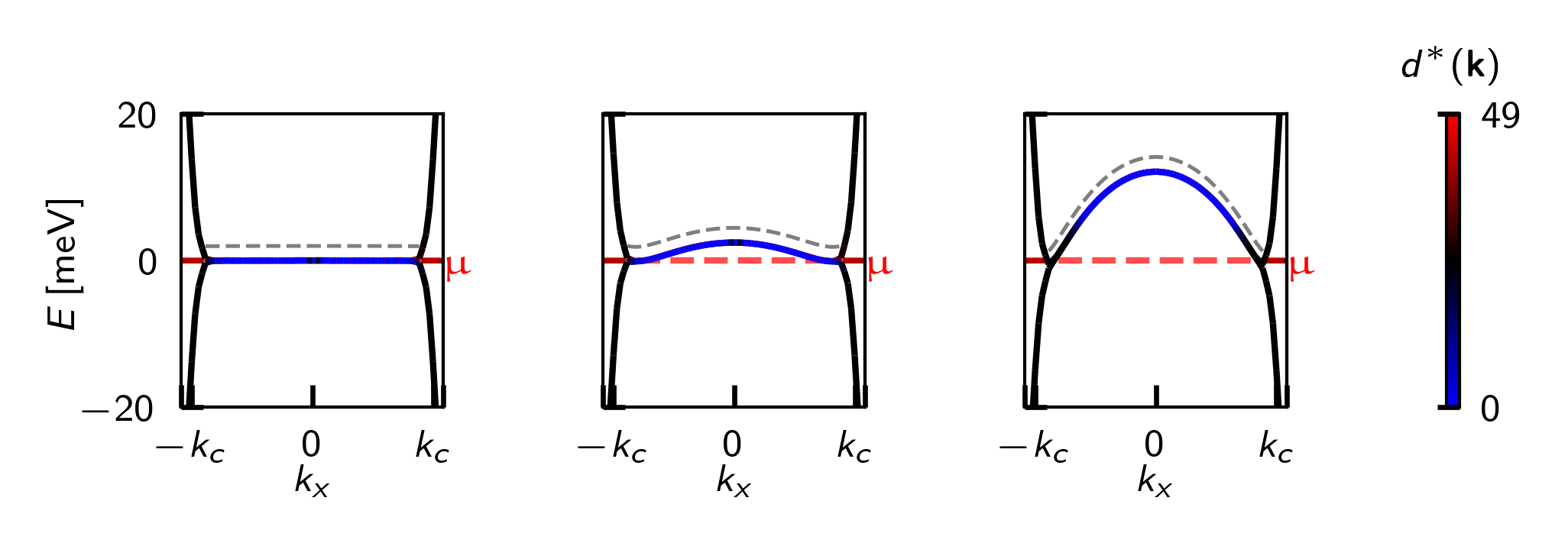}
    \caption{$N=50$ particle--hole-symmetric model.
    (a) Self-consistent band structure at $n=0$, $D=0$. The dashed gray curve shows the analytical dispersion from Eq.~\eqref{eq:supintinduceddisp}, offset by $\SI{2}{meV}$ for clarity.
    (b) Same as (a) at $n=\SI{-5.5e12}{cm^{-2}}=\frac{2}{3}(2n_-)$.
    (c) Same as (a) at $n=\SI{-8.2e12}{cm^{-2}}=2n_-$.}
    \label{fig:suppllphmodelbsfits}
\end{figure*}

\section{Numerical details}
\subsection{Single-particle model parameters}

We use the tight-binding parameters following Ref.~\cite{jungParkTopologicalFlatBands2023}.
Specifically, we take
$v_F=\SI{-547}{meV\,nm}$,
$v_3=\SI{61.66}{meV\,nm}$,
$v_4=\SI{30.3}{meV\,nm}$,
$t_1=\SI{356.1}{meV}$,
and $t_2=\SI{-8.3}{meV}$.
Unless stated otherwise, we do not include additional fixed outer-layer site potentials, because the layer-dependent potentials $U_l$ are determined self-consistently from electrostatics.

For comparison, we have also performed calculations including these additional site potentials used in Ref.~\cite{jungParkTopologicalFlatBands2023}:
$U_{A1}=U_{BN}=0$,
$U_{B1}=U_{AN}=\SI{12.2}{meV}$,
and $U_{AM}=U_{BM}=\SI{-16.4}{meV}$ for $1<M<N$.
The corresponding results are shown in Fig.~\ref{fig:supplvisp}.

\begin{figure*}[t]
    \centering
    \includegraphics[width=0.9\textwidth]{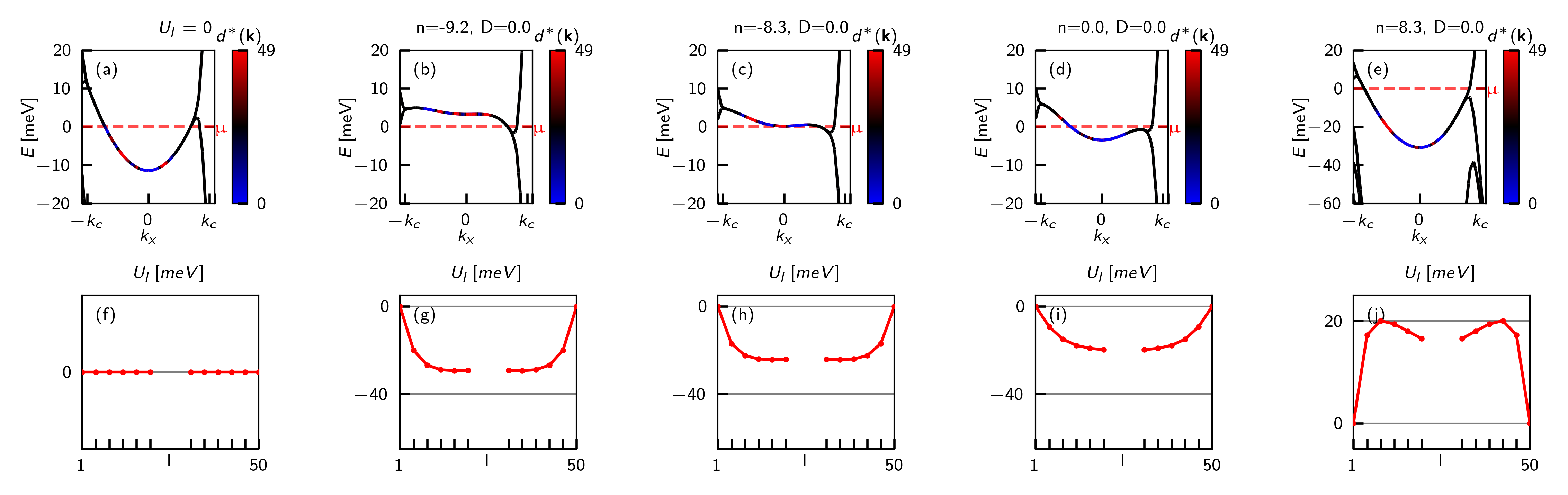}
    \caption{
    Effect of including inversion-symmetric on-site potentials for $N=50$, $D=0$, and $\epsilon_\perp=3$.
    We use the additional site potentials of Ref.~\cite{jungParkTopologicalFlatBands2023}:
    $U_{A1}=U_{BN}=0$,
    $U_{B1}=U_{AN}=\SI{12.2}{meV}$, and
    $U_{AM}=U_{BM}=\SI{-16.4}{meV}$ for $1<M<N$.
    They enter Eq.~\eqref{eq:hamrhombo} as an additive diagonal term
    $\hat{H}_{\mathrm{ISP}}=\operatorname{diag}\!\left(U_{A1},U_{B1},U_{A2},\ldots,U_{AN},U_{BN}\right)$.
    }
    \label{fig:supplvisp}
\end{figure*}

\subsection{Self-consistent calculation}

We obtain self-consistent band structures by iterating Eqs.~\eqref{eq:hamrhombo} and \eqref{eq:potdifference} to convergence.
Here $\dens_l$ denotes the excess charge density on layer $l$ measured relative to charge neutrality. We retain all bands in Eq.~\eqref{eq:hamrhombo} and impose a finite in-plane momentum cutoff on the numerical $\mathbf{k}$ grid.

To aid convergence we use the optimal damping algorithm of Ref.~\cite{lebrisCancesCanWeOutperform2000}.
At $D=0$ we additionally impose $U_1=U_N$, which stabilizes the iteration by enforcing inversion-symmetric boundary conditions.
We restrict to $C_{3z}$-symmetric solutions and therefore work on a reduced $\mathbf{k}$ grid in which points related by $C_{3z}$ are identified.
In practice we use a grid of 519 $\mathbf{k}$ points centered at $\mathbf{k}=0$ with a radial cutoff
$|\mathbf{k}_{\max}|=\SI{0.928}{nm^{-1}}$.
This cutoff exceeds the maximal extent of the (chiral-model) surface band,
$|\mathbf{k}_c|=t_1/|v_F|=\SI{0.65}{nm^{-1}}$,
so the grid covers the entire surface-band region of interest.

Unless otherwise stated, we take the out-of-plane dielectric constant to be $\epsilon_\perp=3$,
motivated by the experimentally determined value in Ref.~\cite{Slizovskiy2021}.

For the DOS plots, since the calculations are performed at $D=0$, where
the two surface bands are degenerate, we divide the resulting value by $2$ 
to correspond to the DOS of a single surface band.

\section{Additional data}

\subsection{Electrostatic phase diagram for $N=5$}

Figure~\ref{fig:figfour} compares the $N=5$ electrostatic phase diagram obtained from the standard linear-potential approximation (top row) with the fully self-consistent electrostatics (bottom row). For $N=5$, the two approaches give very similar results: nonlinear screening effects are modest, and the linear profile provides a reasonable description of the DOS structure and effective mass across the relevant $(n,D)$ range.

\begin{figure*}[t]
    \centering
    \includegraphics[width=0.9\textwidth]{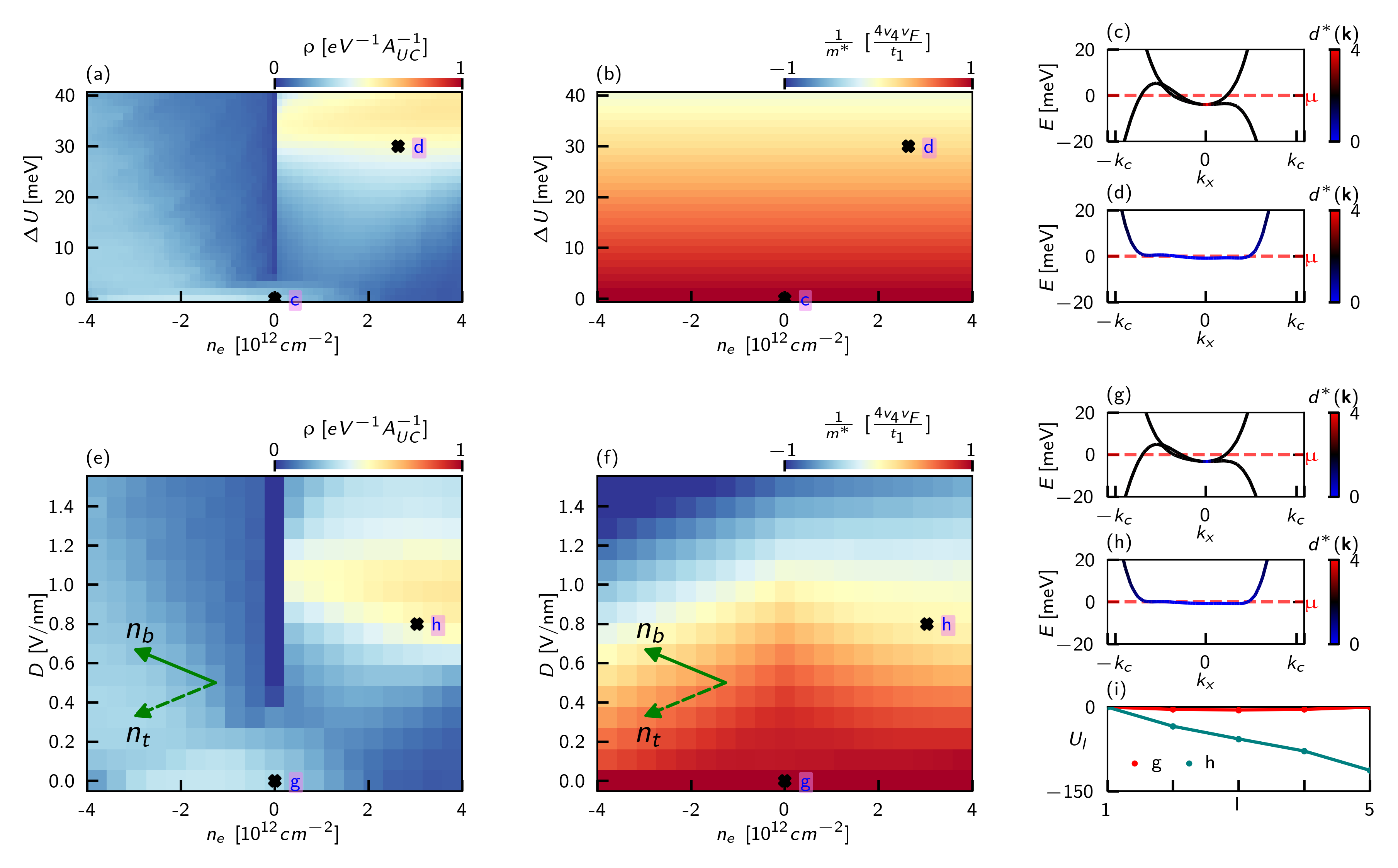}
    \caption{$N=5$ comparison of the linear-potential approximation and the fully self-consistent electrostatics.
    (a) DOS in the $(n,\Delta U)$ plane for a model with a fixed interlayer drop $U_{l+1}-U_l=-\Delta U$.
    (b) Corresponding inverse effective mass (in units of the mass defined in Eq.~\eqref{eq:zeropotdispersion}).
    (c) Band structure at $\Delta U=0$.
    (d) Band structure at large $\Delta U$, focusing on the high-DOS region.
    (e) DOS in the $(n,D)$ plane for the fully self-consistent calculation with potentials determined from Eq.~\eqref{eq:potdifference}.
    (f) Corresponding inverse effective mass.
    (g) Self-consistent band structure in the coexistence region.
    (h) Self-consistent band structure at high $D$ (one surface depleted).
    (i) Self-consistent layer potential profile $U_l$ for the cases shown in (g) and (h).}
    \label{fig:figfour}
\end{figure*}

\subsection{Interaction strength dependence}

Figures~\ref{fig:supplepsone}--\ref{fig:supplepstwelve} illustrate how the self-consistent electrostatic profile and the resulting surface-band dispersion evolve with the out-of-plane dielectric constant $\epsilon_\perp$. For each $\epsilon_\perp$ we show self-consistent results at $D=0$ as a function of carrier density $n$: the band structure (surface band), the corresponding density of states (DOS), and the layer-potential profile (or potential drops). The main trend is monotonic: as interactions are strengthened (decreasing $\epsilon_\perp$), the self-consistent potentials become more surface-localized and more effective at compensating the bare dispersion, producing a systematically flatter surface band and a larger low-energy DOS range. The strongest flattening occurs on the hole-doped side near depletion, where the near-surface potential drop is maximally concentrated and pulls finite-$k$ states down toward $\mathbf{k}=0$.

\begin{figure*}[t]
    \centering
    \includegraphics[width=0.9\textwidth]{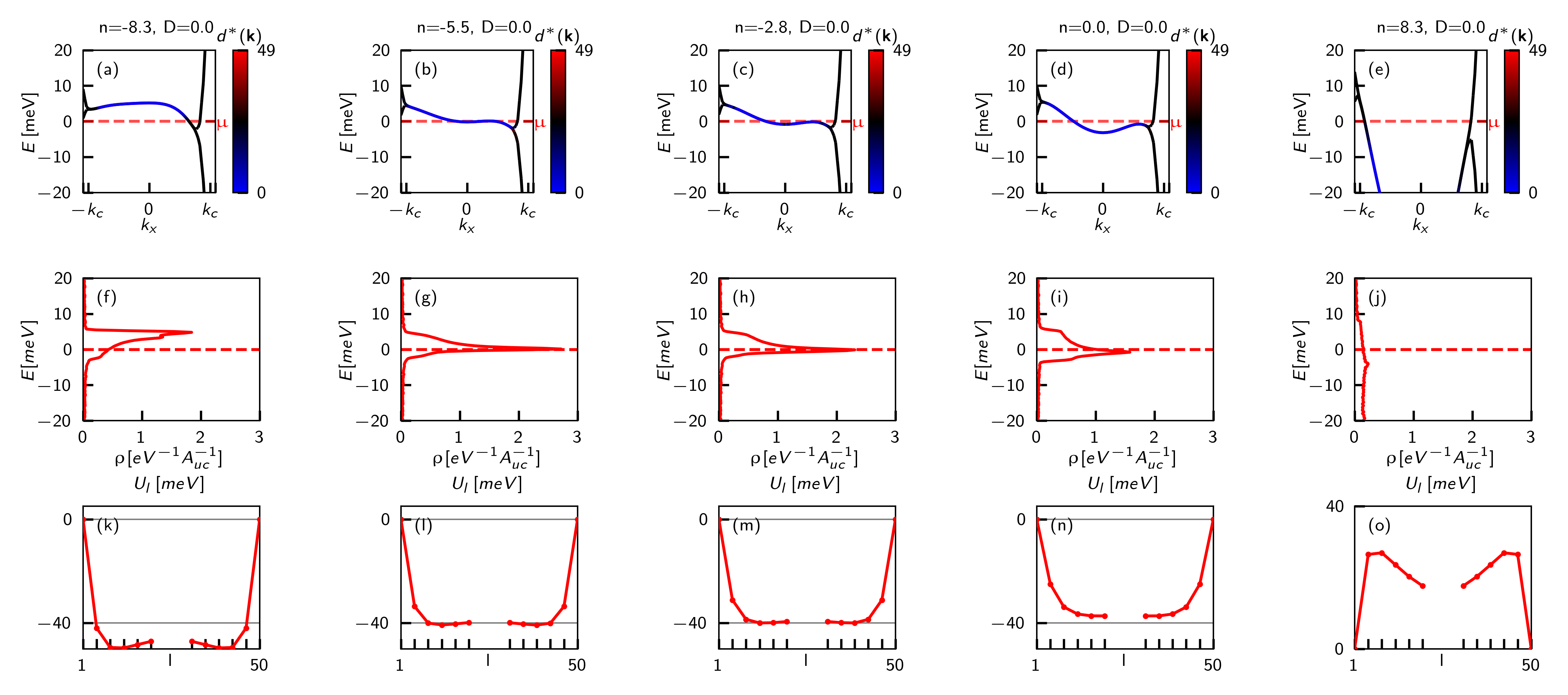}
    \caption{$N=50$, $D=0$: self-consistent surface-band dispersion, DOS, and layer-potential profile for an (unphysically) strong out-of-plane interaction, $\epsilon_\perp=1$, shown for several carrier densities $n$ (in units of $\si{cm^{-2}}$). This extreme limit illustrates the strong-coupling trend: the potential drop becomes highly concentrated near the surface and the surface band is driven toward maximal flattening over the largest density range.}
    \label{fig:supplepsone}
\end{figure*}
\begin{figure*}[t]
    \centering
    \includegraphics[width=0.9\textwidth]{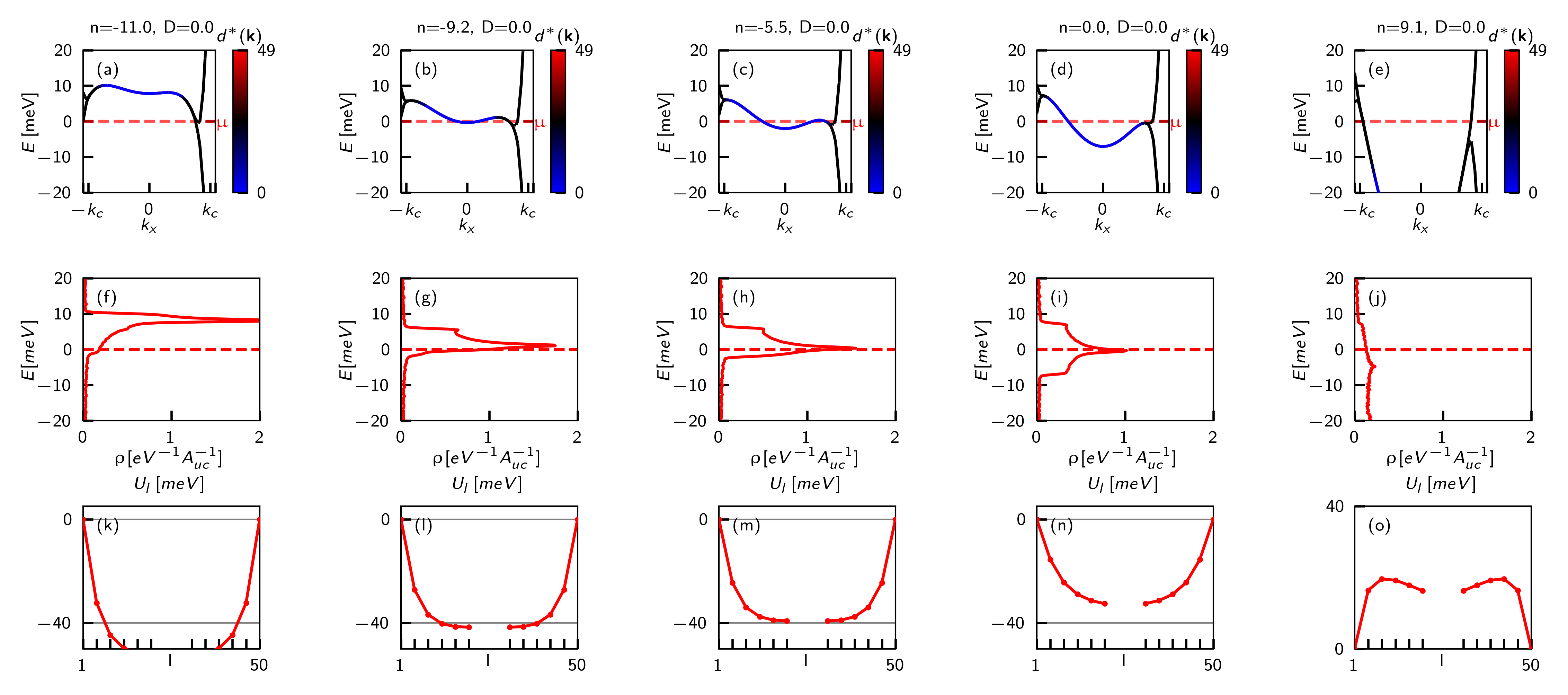}
    \caption{$N=50$, $D=0$: self-consistent surface-band dispersion, DOS, and layer-potential profile at the realistic dielectric constant $\epsilon_\perp=3$ (which is used in the main text). Compared to the $U_l=0$ case, the self-consistent near-surface potential lowers finite-$k$ states and produces substantial band flattening already at charge neutrality, with the strongest flattening on the hole-doped side near depletion.}
    \label{fig:supplepsthree}
\end{figure*}
\begin{figure*}[t]
    \centering
    \includegraphics[width=0.9\textwidth]{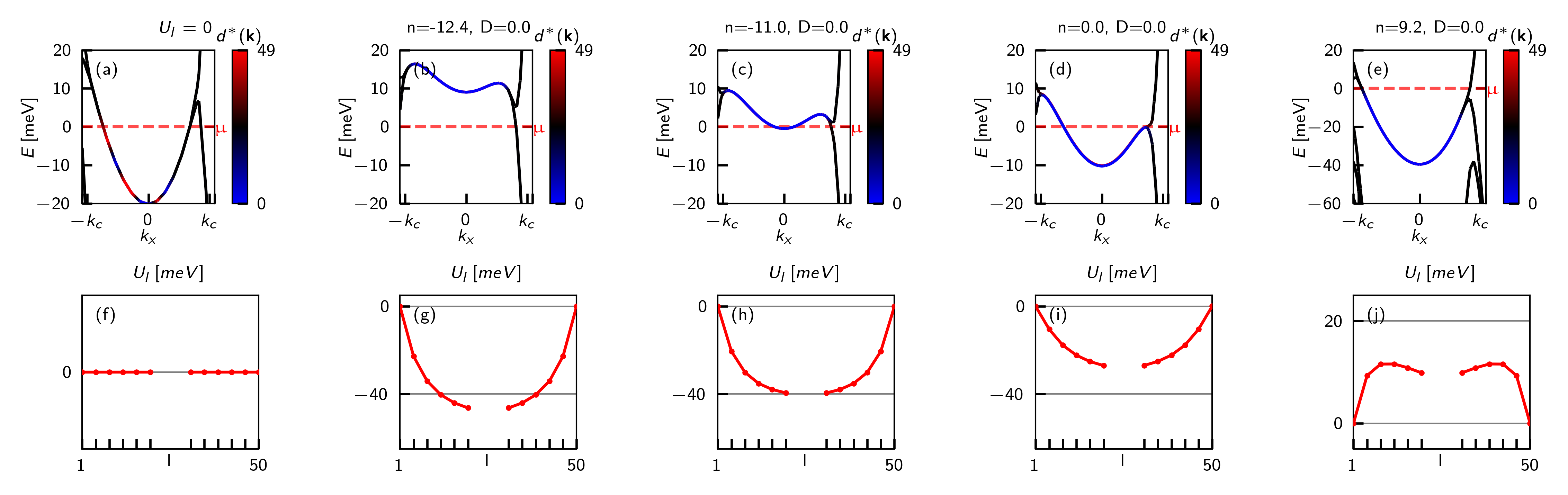}
    \caption{$N=50$, $D=0$: self-consistent surface-band dispersion and layer-potential profile at $\epsilon_\perp=6$ (weaker out-of-plane interactions). The self-consistent potentials are less sharply localized near the surface, and correspondingly the reduction of surface-band bandwidth is weaker than for $\epsilon_\perp=3$.}
    \label{fig:supplepssix}
\end{figure*}
\begin{figure*}[t]
    \centering
    \includegraphics[width=0.9\textwidth]{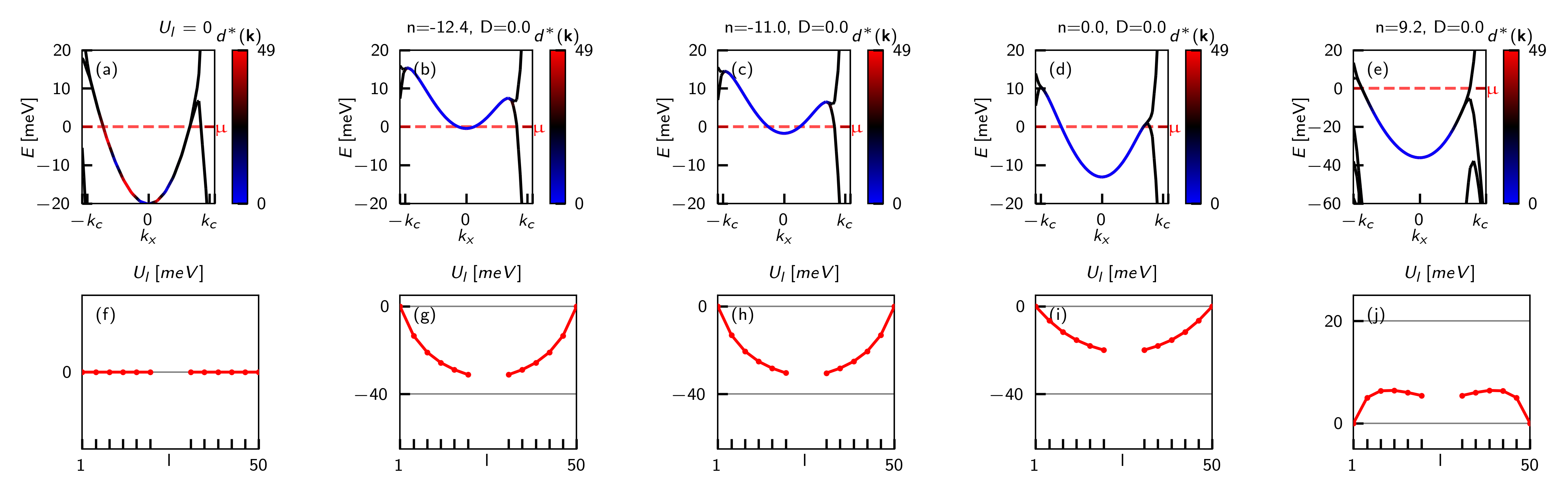}
    \caption{$N=50$, $D=0$: self-consistent surface-band dispersion and layer-potential profile at $\epsilon_\perp=12$ (very weak out-of-plane interactions). In this regime the induced potentials are small and spread over more layers, so the surface band remains comparatively more dispersive and the DOS enhancement is reduced.}
    \label{fig:supplepstwelve}
\end{figure*}

\end{widetext}
\end{document}